\documentclass[a4paper,fleqn,usenatbib]{mnras}

\usepackage{newtxtext,newtxmath}

\usepackage[T1]{fontenc}
\usepackage{ae,aecompl}

\usepackage{graphicx}	
\usepackage{amsmath}	
\usepackage{amssymb}	

\usepackage{placeins}
\usepackage{natbib}
\usepackage{hyperref}
\usepackage{xcolor}
\usepackage{verbatim}
\usepackage{listings}
\usepackage{hyperref}
\usepackage{cleveref}
\usepackage{mathrsfs}
\usepackage{dsfont}
\usepackage{enumitem}

\title[Separating diffuse from point-like sources]{Separating diffuse from point-like sources -  a Bayesian approach}

\author[J. Knollm\"uller et al.]{
	Jakob Knollm\"uller,$^{1,2}$\thanks{E-mail: jakob@mpa-garching.mpg.de}
	Philipp Frank,$^{1,2}$
	Torsten A. En\ss lin$^{1,2}$
	\\
	$^{1}$Max Planck Institute for Astrophysics, Karl-Schwarzschild-Str. 1, 85741 Garching, Germany\\
	$^{2}$Ludwig-Maximilians-Universit\"at M\"unchen, Geschwister-Scholl-Platz{\small{}~}1,
	80539 Munich, Germany\\
}

\date{Accepted XXX. Received YYY; in original form ZZZ}

\pubyear{2015}

\begin{document}
\label{firstpage}
\pagerange{\pageref{firstpage}--\pageref{lastpage}}
\maketitle

\begin{abstract}
We present the \texttt{starblade} algorithm, a method to separate superimposed point sources from auto-correlated, diffuse flux using a Bayesian model. Point sources are assumed to be independent from each other and to follow a power-law brightness distribution. The diffuse emission is described as a non-parametric log-normal model with a priori unknown correlation structure. This model enforces positivity of the underlying emission and allows for variation in the order of magnitudes. The correlation structure is recovered non-parametrically in addition to the diffuse flux and is used for the separation of the point sources. Additionally many measurement artifacts appear as point-like or quasi-point-like effects, not compatible with superimposed diffuse emission. An estimate of the separation uncertainty can be provided as well. We demonstrate the capabilities of the derived method on synthetic data and data obtained by the Hubble Space Telescope, emphasizing its effect on instrumental artifacts as well as physical sources. The performance of this method is compared to the background estimation of the SExtractor method, as well as to a denoising auto-encoder.
\end{abstract}




\section{Introduction}
Our Universe overwhelms us with richness of structure and complexity. In order to grasp its governing processes, one has to focus on one single aspect. Unfortunately most observations are sensitive to a large variety of phenomena. Their accurate separation into distinct components is critical, as it will influence any further analysis. In this paper we want to deal with the problem of separating point sources from diffuse emission. We develop the \texttt{\textbf{starblade}} (\textbf{st}ar and \textbf{a}rtifact \textbf{r}emoval with a \textbf{b}ayesian variationa\textbf{l} \textbf{a}lgorithm from \textbf{d}iffuse \textbf{e}mission) method to separate those two classes of structures occurring in astronomical imaging. Point-like sources can be extremely bright as well as extremely faint, therefore they inhabit a huge dynamic range. By definition they are too small to be spatially resolved and are rather independent of their apparent surroundings. Diffuse, extended emission can be spatially resolved. Large structures are almost impossible to observe without being affected by superimposed point sources. In other contexts weak point-like structures embedded in a diffuse background are of interest. Here it might be important not to be blinded by the background emission.

Another component which is present in real observation are artifacts originating from the measurement process itself.
Those artifacts can exhibit point-like characteristics, such as cosmic ray hits on the detector or edges. They are often correlated only along one image direction and relatively unrelated to the distant cosmos.

Conceptually the two components, point sources and diffuse emission, are independent, therefore an independent component analysis (ICA) \citep{ICAAA} should be the method of choice in order to separate them. ICA separates stochastically independent components by their different appearance.  It is problematic for classical ICA algorithms to have fewer data channels than components. Considering only one individual image the separation is an ill-posed problem. In principle any flux could be explained by either only point sources or diffuse emission, but both scenarios are neither plausible nor useful. In order to separate the components one needs to add additional information, judging a possible separation by some criteria. There will not be a unique solution to this problem, but a large number of plausible separations. An answer to the question of the separation can therefore be only of probabilistic nature. The likelihood of one configuration derives from a prior probabilities, which manifest some knowledge on the components. For this we have to mathematically formulate the concept of diffuse emission and point-like sources and then confront them to the data in order to obtain a plausible separation.

In this paper we present a method how to separate those two components from a single image. We will use physically motivated models for the description of point-like and diffuse components and derive a posterior estimate of the separation using Bayes' theorem. As the posterior is not accessible analytically, we perform a variational approximation to the posterior quantities, which is capable of capturing uncertainty. The resulting algorithm will be a non-parametric,  hand-crafted ICA method specifically tailored to separate diffuse from point-like sources. It can also provide an estimate of the separation uncertainty at every position. We will use the formalism of information field theory (IFT) \citep{IFT}, which allows us to easily generalize the methods to additional spatial dimensions or resolutions.

This paper deals with the pure diffuse-point source separation problem. Complications due to the imperfection of the data originating from noise or point spread functions are ignored. There are several reasons for such an approach. First, there are data sets that are indeed of such high fidelity that the assumption of vanishing noise is basically fulfilled. Second, for moderate fidelity data a detailed noise modeling might be too expensive, given the scientific focus at hand. Third, the method is useful to detect and remove point-like artifacts from images, as e.g. generated by cosmic ray hits on CCDs of space based telescopes. Additionally, for the imaging of low-fidelity data, the separation of point-like and diffuse flux given a perfectly assumed sky brightness is a useful internal step of the denoising and imaging algorithm, as we will explain in Appendix \ref{ap:lager_picture}. We will explore the algorithms capability to generalize to such imperfect situations by confronting it with real data in on of our examples.

The traditional approach how to deal with distracting point sources is to mask them out. A point source is identified by some criterion and its area is removed from the image. This approach has two disadvantages. First, the masking might effect further analysis if it is not carefully considered and therefore could corrupt results. Secondly, it is hard to identify and properly mask weak point sources. In order to identify them it is vital to consider the surrounding area and its correlation structure. 

A popular method to extract point sources in images is the SExtractor software \citep{sextractor}. It removes background, identifies sources, classifies them, extracts characteristic features and builds catalogs. Another widely used software is DAOPHOT \citep{stetson1987daophot}. It specializes in crowded fields. In both methods the background removal is done by a heuristic scheme, which for many applications performs excellent, especially if the background can be approximated to be constant and sources are sparse. A Bayesian version, which also provides uncertainties on those quantities is the Background-Source separation method  \citep{guglielmetti2009method}. 
Another method which is comparable to the one presented here is \citet{popowicz2015method}, which relies on local neighborhoods and a morphological distance transform, but is not derived from probabilistic principles. 

The problem of separating point-sources and diffuse emission can also be regarded as recovering a diffuse component, which is corrupted by point-sources. The recent development in deep learning lead to a large variety of different architectures, which are capable to learn such tasks, based on huge amounts of data. One such architecture is the denoising auto-encoder (DAE)  \citep{DAE}. It is trained on pairs of corrupted images and its ground truth. Here the corrupted images are typically generated artificially to mimic some kind of degradation, such as Gaussian or Salt-and-Pepper noise \citep{xie2012image}. In analogy to this we will compare our method with a denoising auto-encoder trained on pairs of diffuse emission and a point-source corrupted version.

Let us briefly outline the structure of this paper. We will start in Sec. \ref{sec:datamodel} with introducing the underlying description of the data, followed by a discussion of point-like and diffuse emission in Sec. \ref{sec:pointlike} and Sec. \ref{sec:diffuse}, respectively. The full mathematical structure of the problem is derived in Sec. \ref{sec:full_picture}. Solving the problem requires some further numerical considerations, which are outlined in Sec. \ref{sec:numerical}. The variational approach we use to infer an approximated posterior separation is described in Sec. \ref{sec:variational}, followed by a brief summary of the algorithmic steps of the \texttt{starblade} algorithm in Sec. \ref{sec:algorithm}. 
We validate our algorithm by applying it to synthetically generated data, and demonstrate its application to real data, an image of the $\mathrm{M}100$ galaxy, obtained by the Hubble Space Telescope in Sec. \ref{sec:examples}. In both cases we compare its performance with the background estimation step of the SExtractor algorithm and an denoising auto-encoder (DAE).
We conclude in Sec. \ref{sec:conclusion}.  How the here presented method can be used in larger inference frameworks is outlined in Appendix. \ref{ap:lager_picture}. In Appendix \ref{ap:DAE} we describe in detail the implementation, architecture and training of the DAE.

\section{The data model}
\label{sec:datamodel}
The data we are considering consists of a superposition of two components. On the one side spatially correlated, positive diffuse flux, on the other side spatially uncorrelated, also positive, point-like flux. Negative flux values are unphysical and we will  exclude them by enforcing the positivity of the components. To this end we express them in terms of their logarithmic brightness.  
\begin{align}
\label{eq:data}
d = e^{s} + e^{u}
\end{align}
The logarithmic diffuse emission is expressed in $s$, the logarithmic point-like flux in $u$. The quantities $s$ and $u$ are fields, meaning they are functions of the location $x$. The exponential function in Eq. \ref{eq:data} and other functions are applied point-wise in IFT, meaning $(e^{s})_x = e^{s{_x}}$.

We will approach the separation problem from the probabilistic perspective and we can use the data equation  Eq.\ref{eq:data} to derive the likelihood of the data, given the point sources and diffuse emission. As the data is not exposed to any randomness for given pairs of $s$ and $u$ in the noiseless limit, this likelihood is expressed by a delta distribution. 
\begin{align}
\label{eq:delta}
\mathcal{P}(d \vert s, u) = \delta(d - e^{s} - e^{u})
\end{align}
We can combine this likelihood with a prior that models what we mean by point-like and diffuse emissions, allowing their separation. The separation is done by applying Bayes theorem, 
\begin{align}
\label{eq:bayes}
\mathcal{P}(s,u\vert d) = \frac{\mathcal{P}(d\vert s, u) \mathcal{P}(s) \mathcal{P}(u)}{\mathcal{P}(d)} 
\end{align}
and asking for the most plausible a posterior separation of $d$ into $e^s$ and $e^u$.
Note that we assumed point and diffuse sources to be independent of each other, which implements the fundamental assumption of an ICA:
\begin{align}
\mathcal{P}(s,u) = \mathcal{P}(s) \mathcal{P}(u) \text{.}
\end{align}
We now need expressions for the prior distributions $\mathcal{P}(u)$ and $\mathcal{P}(s)$, defining the characteristics of point-like and diffuse emission, respectively.

\section{Point-like emission}
\label{sec:pointlike}

The defining features of point sources is their spatial independence and strong diversity in brightness.
The independence is expressed by their joint probability distributions factorizing into independent probabilities for each position.
\begin{align}
\mathcal{P}(u) = \prod_x \mathcal{P}(u_x)
\end{align}
The brightness distribution of the individual point sources can often be argued to follow a power-law, as we expect the number of sources to scale with the observed volume and the brightness to decrease with distance. In an Euclidean universe with uniformly distributed point sources the exponent of this distribution is expected to be $\alpha = 1.5$. A detailed discussion of the choice of this parameter can be found in \citet{D3PO} and also in \citet{guglielmetti2009method}. In practical applications this value might be too restrictive, as the universe is not Euclidean and the sources exhibit an evolution with cosmic time, which translates to a distance dependence. Thus other values for $\alpha$ might be chosen. The choice of this parameter will influence the separation and it will define the sensitivity of the method in either the direction of assigning more flux to the diffuse emission or to point sources.
It is important to note that the impact of $\alpha$ in general is not scale independent. An increase or decrease in resolution splits or merges pixels and the point sources associated with them. This splitting or merging of point sources changes in general the effective brightness distribution.  Only for $\alpha = 1.5$, a change in resolution  has no effect. Any other value of $\alpha$ expresses a power law brightness distribution only for the chosen resolution exactly. Changing the resolution without readjusting $\alpha$ actually means to chose a different brightness distribution. For the brightest sources, this subtlety does not make a big difference. A discussion of this matter can as well be found in \citet{D3PO}.

In order to ensure the normalization of the prior distribution for any choice of $\alpha$ and for numerical reasons we introduce a low-brightness cut-off. It will suppress vanishing brightness values, stabilizing the algorithm. A physical motivation to this cut-off is the finite extension of our host galaxy, the Milky Way and the finite extent of the look back light cone in the universe. After a certain distance we do not expect a large number of point sources. This leads to the choice of an inverse gamma distribution for the point sources, which reads:
\begin{align}
\mathcal{P}(u) = \mathcal{I}(e^u, \alpha, q) = \frac{q^{\alpha-1}}{\Gamma(\alpha-1)}  e^{-(\alpha-1)^\dagger u } e^{-q^\dagger e^{-u}} \text{.}
\end{align}
The $\dagger$ expresses the complex conjugated, transposed vector or field. 
We will set $q$ to small values in order not to influence the separation in a significant way.
\section{Diffuse emission}
\label{sec:diffuse}
For the diffuse emission we propose a non-parametric log-normal model, which assumes the logarithmic flux to be Gaussian distributed. The spatial correlations are expressed in the correlation structure of this Gaussian distribution. 
\begin{align}
\mathcal{P}(s) = \mathcal{G}(s,S) \equiv \frac{1}{\vert 2\pi S\vert^{\frac{1}{2}}} 
e^{-\frac{1}{2}s^\dagger S^{-1} s}
\end{align}
The correlation structure $S$ is a priori unknown. Assuming prior homogeneity and isotropy, it is represented by a diagonal operation in the Fourier space, according to the Wiener-Khintchin theorem \citep{Wiener,Khintchin}, and is described by a one dimensional power spectrum. With these assumptions we can express the correlation structure $S$ compactly in terms of:
\begin{align}
	S = \mathbb{F}^\dagger \widehat{\left( \mathbb{P} e^\tau\right)} \mathbb{F}
\end{align}
We express the power spectrum in term of its logarithmic power spectrum $\tau$ to ensure positivity. The isotropy operator $\mathbb{P}$ distributes this one dimensional power spectrum into the full harmonic space. The $\widehat{}$ indicates the raising of this field to an diagonal operator, and finally the Fourier transformations $\mathbb{F}$ implement the homogeneity assumption. We parametrize the prior correlation structure in terms of its logarithmic power spectrum $\tau$. The inference of this parameter will be part of our overall procedure. This corresponds to the critical filter, which is derived in detail in \citet{ensslinfrommert} and \citet{smoothpower}.

 Overall we describe the diffuse emission  by a log-normal model with unknown a priori correlation structure. This model has been applied in an astrophysical context to describe diffuse structures in various situations \citep{D3PO,RESOLVE, QPO, NCF}.
 
\section{The full picture}
\label{sec:full_picture}

Now we have all prior distributions in order to calculate the posterior for the diffuse and point-like flux, as described in Eq. \ref{eq:bayes}. We can get rid of one quantity by marginalizing out the delta distribution from the likelihood contribution Eq. \ref{eq:delta}. We choose to perform the marginalization over $s$.
\begin{align}
\mathcal{P}(u\vert d) &= \int \mathcal{D}s \frac{\mathcal{P}(d\vert s,u)\mathcal{P}(s)\mathcal{P}(u)}{\mathcal{P}(d)}\\
&= \frac{\mathcal{P}(u)}{\mathcal{P}(d)} \int \mathcal{D}s \: \delta(d-e^s -e^u) \: \mathcal{G}(s,S) \\
&= \frac{\mathcal{P}(u)}{\mathcal{P}(d)} \mathcal{G}(\mathrm{ln}(d-e^u),S) \frac{1}{\prod _x\vert d_x-e^{u_x} \vert}
\end{align}
All terms not containing any dependence on $s$ can be pulled out of the integral. Performing the integral replaces $s$ in its Gaussian prior with $\mathrm{ln}(d-e^u)$ to fulfill the constraint. In addition we get the factor $\prod_x\vert d_x-e^{u_x}\vert^{-1}$ originating from the change in variables in order to perform the integral.
The resulting expression only depends on the logarithmic diffuse flux $u$. For mathematical convenience we investigate
\begin{align}
\label{eq:hamiltonian}
\mathcal{H}(u\vert d) \equiv &  - \mathrm{ln} \: \mathcal{P}(u\vert d)  \\
=& \: \mathcal{H}_0
+ \frac{1}{2}\mathrm{ln}(d -e^u)^\dagger S^{-1}  \mathrm{ln}(d -e^u)  \nonumber \\
&+ (\alpha - 1)^\dagger u + q^\dagger e^{-u} 
+ 1^\dagger \mathrm{ln}(d-e^u) \text{.}
\end{align}
The expression above fully describes the problem. It corresponds to the negative log-posterior, or, in the language of IFT, the information Hamiltonian.

\section{Numerical considerations}
\label{sec:numerical}

Our inference will be based on the minimization of some target functional with respect to some parameters. In the current formulation of the setup we have numerically problematic expressions of the form $\mathrm{ln}(d-e^u)$, which can be temporarily ill-defined during the inference calculations due to negative values within the logarithm. We can overcome this limitation by introducing a separation field $a$, which ranges in each pixel within $[0,1]$, attributing a fraction $a$ of its image value $d$ to the point source $ e^u \equiv ad$, and the fraction $e^s =(1-a)d$ to diffuse emission.

In order to do so we introduce the additional constraint
\begin{align}
\mathcal{P}(u\vert ad) = \delta(u - \mathrm{ln}(ad)) \text{,}
\end{align}
which allows us to reformulate the problem Hamiltonian in terms of $a$ via marginalization over $u$.
\begin{align}
\mathcal{H}(a\vert d)  =& \: \mathcal{H}_0 
+ \frac{1}{2}\mathrm{ln}((1-a)d)^\dagger S^{-1}  \mathrm{ln}((1-a)d)  \nonumber \\
&+ (\alpha - 1)^\dagger \mathrm{ln}(ad) + q^\dagger \frac{1}{ad} 
 - 1^\dagger \mathrm{ln}((1-a)d)
 + 1^\dagger \mathrm{ln}(a) 
\end{align}
The last term originates from the functional determinant of the substitution.
To ensure that the separation field ranges between zero and one, we parametrize it with a sigmoid function applied to some underlying field $b$. A function fulfilling a sigmoid shape, ranging from $0$ to $1$ is 
\begin{align}
\label{eq:sigmoid}
a = \frac{1}{2} (\mathrm{tanh}(b)+1) \text{.}
\end{align}
Finally, this internal separation field $b$ will be the quantity we try to infer in order to separate point sources from diffuse emission. Again, we can introduce it to the model formulation via an additional probability distribution $\mathcal{P}(a\vert b)$ on $a$, which, when marginalized out, replaces every $a$ with the expression above. Another functional determinant adds through this substitution.

The sigmoid function given in Eq. \ref{eq:sigmoid} approaches for large absolute values of $b$ its respective boundary of $0$ or $1$ exponentially. Therefore at some point increasing values of  $\vert b\vert $ do not change the separation in any significant way. If only one component is present at one location, there is no resistance for the algorithm to push the value of $b$ to arbitrarily high values. This can cause numerical instabilities, as it represents unconstrained degrees of freedom within the problem. To counteract this behavior we will introduce an additional weak Gaussian prior on  $b$, centered at zero. Values of $ \vert b \vert > 10$ impose a dynamical range of the ratio between the two components of roughly $1 : 10^{9} $. We want to keep the values of $b$ within a range to explain any separation between point source and diffuse flux, but regularizing against an unnecessary drift. For this we add a small, quadratic prior energy for $b$. The full description of the problem is then expressed in the Hamiltonian
\begin{align}
\label{eq:full_hamiltonian}
\mathcal{H}(b\vert d)  =& \: \mathcal{H}_0 
+ \frac{1}{2}\mathrm{ln}((1-a)d)^\dagger S^{-1}  \mathrm{ln}((1-a)d) \nonumber  \\
&+ (\alpha - 1)^\dagger \mathrm{ln}(ad) + q^\dagger \frac{1}{ad} + \frac{1}{2 \sigma^2} b^\dagger b \nonumber \\
&- 1^\dagger \mathrm{ln}((1-a)d) + 1^\dagger \mathrm{ln}(a) - 1^\dagger \mathrm{ln}(1-\mathrm{tanh}^2(b)) \text{.}
\end{align}
Again the last term originates from the functional determinant of the final substitution.
The free parameters of this model are the correlation structure $S$, the cutoff of the brightness distribution $q$ and its scaling  behavior $\alpha$, and  the prior standard deviation $\sigma$ of the $b$ field, which is chosen large, for example $\sigma = 3$, so that it usually has a small effect, which mainly restricts values the values of $x$ between roughly $-10$ and $10$, providing almost the full range of the separation field $a$ between $0$ and $1$.

We propose to set a low value to the cutoff parameter to minimize its impact to the inference, say $q = 10^{-10}$ for a flux scale in the vicinity of unity. In cases one has reasons to assume that the number of  faint sources is suppressed, it can be adjusted accordingly.

The only parameter we cannot fix generally is the value of the scaling parameter $\alpha$, which  influences the outcome of the separation. The larger its value, the stronger point sources are suppressed, the more the flux will be attributed to the diffuse emission, and vice versa. This effect is most sensitive to regions of superimposed fluxes. It determines how significantly point sources have to stick out, in order not to be considered part of the diffuse flux. This parameter will have to be set by the user, depending on the questions asked to the data. A lower limit to  $\alpha$  is the value $1$, which corresponds to an uninformative prior on the scale. It has to be larger than one for the prior distribution to be normalizable. Small $\alpha$ correspond to high point-source flux. Choosing such a value makes it easy for the algorithm to explain the flux with the point-like component, suppressing the small scales of the diffuse component. Choosing a large $\alpha$, all flux will end up in the diffuse component, as any point source component is suppressed strongly. For values of $\alpha$ in between those extremes, separations are achieved which balance diffuse emission and point sources according to this parameters. If no specific reason is given to choose $\alpha$, we recommend the resolution independent choice of $\alpha=1.5$, also favored in a homogeneous Euclidean  Universe.

\section{Variational inference}
\label{sec:variational}
We do not have access to the posterior distribution as the normalization is not tractable, so we will rely on a variational scheme to obtain posterior estimates of the separation. This approach is more robust against the choice of inappropriate hyper prior parameters, compared to the popular maximum posterior estimate (MAP). We will demonstrate this in one of our examples. The inference of the variational parameters is done by minimizing the Kullback-Leibler divergence \citep{KLdivergence} between the true posterior $\mathcal{P}(b\vert d)$ and an simpler, approximative posterior $\widetilde{\mathcal{P}}(b\vert d)$, which is given by

\begin{align}
\label{eq:KLDivergence}
\mathcal{D}_{KL}(\widetilde{\mathcal{P}}(b\vert d) \vert \vert \mathcal{P}(b \vert d)) &= \int \mathcal{D}b \: {\widetilde{\mathcal{P}}}(b\vert d) \:\mathrm{ln}\:\frac{{\widetilde{\mathcal{P}}}(b\vert d)}{ \mathcal{P}(b \vert d)} \\
&= \langle \mathcal{H}(b\vert d)\rangle_{\widetilde{\mathcal{P}}(b\vert d)} - \langle \widetilde{\mathcal{H}}(b\vert d)\rangle_{\widetilde{\mathcal{P}}(b\vert d)} \text{.}
\end{align}
As an approximate distribution we will use a Gaussian distribution which has a number of convenient properties and it already captures the crucial feature of an uncertainty, therefore the approximation has the form $\widetilde{\mathcal{P}}(b\vert d) = \mathcal{G}(b-\bar{b}, B)$ and it remains to determine the values for $\bar{b}$ and $B$ by minimizing Eq. \ref{eq:KLDivergence}. We can calculate the gradient with respect to $\bar{b}$ using the identity
\begin{align}
\frac{\delta \mathcal{D}_{KL}}{\delta \bar{b}} =\left \langle \frac{\delta H(b\vert d)}{\delta b}\right\rangle_{\mathcal{G}(b-\bar{b}, B)}
\end{align}
and we can solve for $B$ by setting the gradient of the KL divergence with respect to it to zero and solve the resulting equation. As we chose a Gaussian approximation this becomes
\begin{align}
B^{-1} = \frac{\delta^2 \mathcal{D}_{KL}}{\delta \bar{b} \delta \bar{b}^\dagger} \equiv \left \langle \frac{\delta^2 H(b\vert d)}{\delta b \delta b^\dagger}\right\rangle_{\mathcal{G}(b-\bar{b}, B)} \text{.}
\end{align}
This covariance is equally the curvature of the KL with respect to its mean and we will recycle it within our minimization to obtain a Newton scheme. In order to approximate the expectation values we draw a set of independent samples from our approximate distribution and replace the integral over the distribution with a simple sum. More detailed discussions of approximations of this kind can be found in \citet{NICAAC} and \citet{NCF}. Note that we avoid to explicitly represent the covariance at any time. We can extract any desired quantity from it by solving a system of linear equations using numerical schemes, such as the conjugate gradient method \citep{conjugate}. This is necessary as its size scales quadratic with the number of image pixels.

In order to infer the unknown correlation structure we refer to the critical filter described in \citet{ensslinfrommert}, which assumes a priori homogeneity and isotropy to formulate the correlation structure of the diffuse component as power spectrum in the harmonic domain. The previously mentioned samples can be used here as well to ensure the required uncertainty corrections.

\section{The starblade algorithm}
\label{sec:algorithm}
The \texttt{starblade} algorithm minimizes the variational KL divergence between the true posterior distribution and an approximate Gaussian distribution and additionally estimates the prior correlation structure of the diffuse component. It implements the following steps:

\begin{enumerate}[label=\arabic*)]
\item Initialize the logarithmic power spectrum $\tau$ and the internal separation field $b$.
\item Draw a set of samples from the approximate Gaussian distribution at the current position as described in \citet{NICAAC} or \citet{NCF}.
\item Use these samples to obtain a statistical estimate of the KL divergence, gradient and curvature according to Eq. \ref{eq:full_hamiltonian}.
\item Minimize the estimated KL divergence to obtain an improved internal separation field, preferably with a second order Newton scheme.
\item Update the logarithmic power spectrum with the critical filter according to \citet{smoothpower}.
\item Iterate this procedure with updated parameters, starting from the second step until desired convergence is achieved.
\item After convergence a set of approximate posterior samples can be drawn to further investigate the result.
\end{enumerate}

\section{Examples}
\label{sec:examples}

In order to illustrate the  behavior of the \texttt{starblade} method we will show two examples. The first example uses synthetically generated data using a log-normal diffuse component with artificially added point-sources of varying magnitude. We apply the algorithm three times to this data with different choices of $\alpha$ to compare its impact on the separation. Using synthetic data, we do have access to the ground truth, which allows us to evaluate the algorithms fidelity, and compare them to other methods. We will use the same test data and apply two configurations of the background estimation of the  SExtractor method \citep{sextractor}, and additionally we train a denoising convolutional  auto-encoder on exactly this model. 
Additionally we infer the MAP solution for an inappropriate choice of $\alpha$ to demonstrate the robustness of the variational approach compared to MAP.

In the second example we separate an observation of the galaxy $\mathrm{M}100$ by the Hubble Space Telescope. This data does not fully fulfill the initial assumption of a noise free image. Nevertheless we will be able to obtain reasonable results. Here we also apply the other two methods and compare the results. Unfortunately we do not have access to the ground truth within this real data application, so instead we will check the result for its compatibility with theoretically motivated assumptions on independence and signatures of correlated and point-like emission. We also discuss the relation of the methods with respect to each other.

We implemented the algorithm in \texttt{Python}, using the numerical information field theory package \texttt{NIFTy} \citep{Nifty,Nifty3}.

\subsection{Synthetic data}
\label{sec:mock_example}
In the first example we will generate data according to the underlying model and investigate the algorithms behavior. In this scenario we do have access to the ground truth, which allows us to derive the quantitative performance compared to other methods. We will compare the results of the \texttt{starblade} to the background estimation step of the SExtractor method \citep{sextractor}, as well as with the performance of a denoising auto-encoder (DAE) \citep{DAE} trained on the identical model.

\begin{figure}
	\includegraphics[scale=0.6]{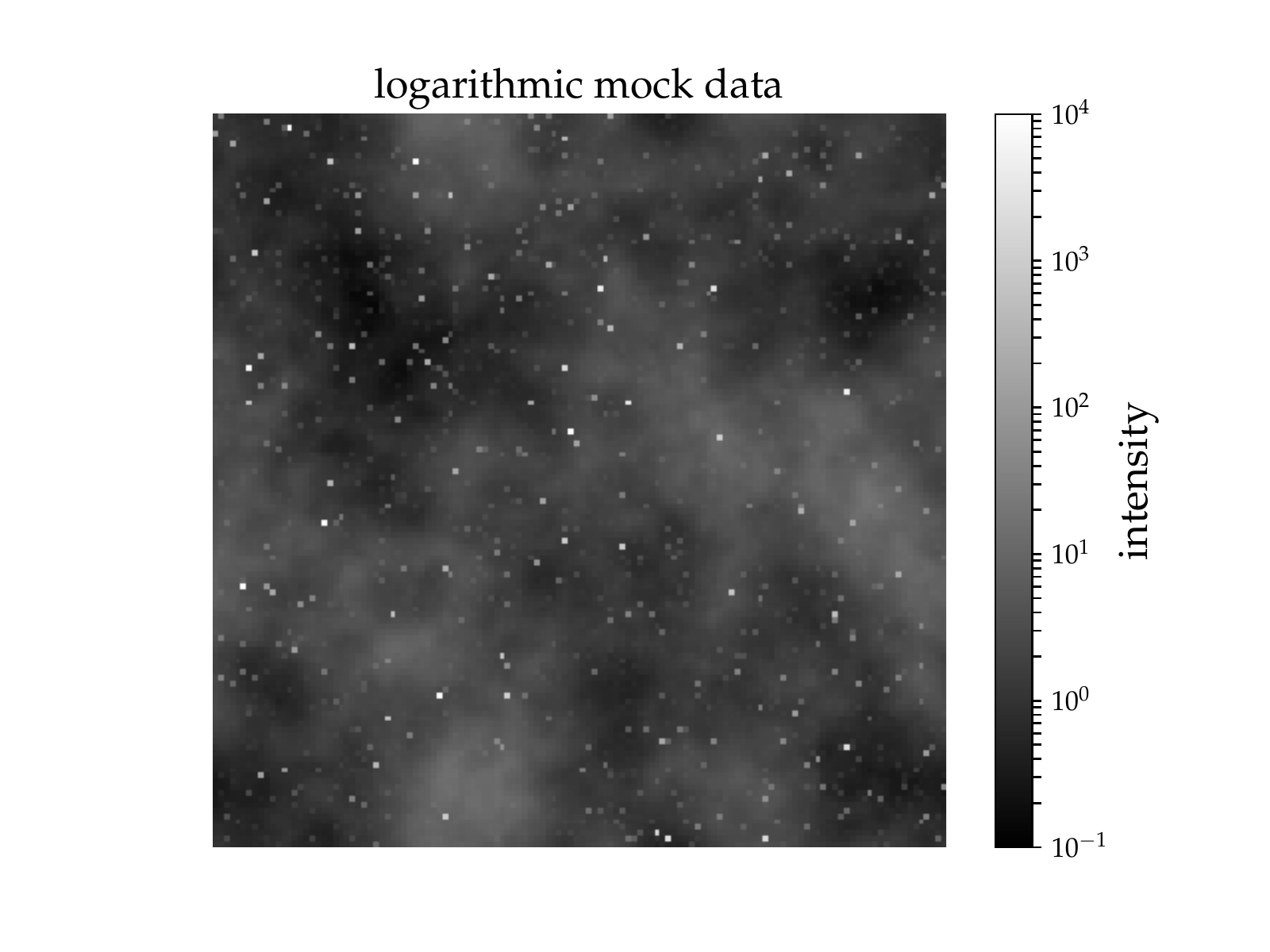}
	\caption{Data generated according the proposed model on an logarithmic scale.}
	\label{fig:2d_data}
\end{figure}

\begin{figure}
	\includegraphics[scale=0.45]{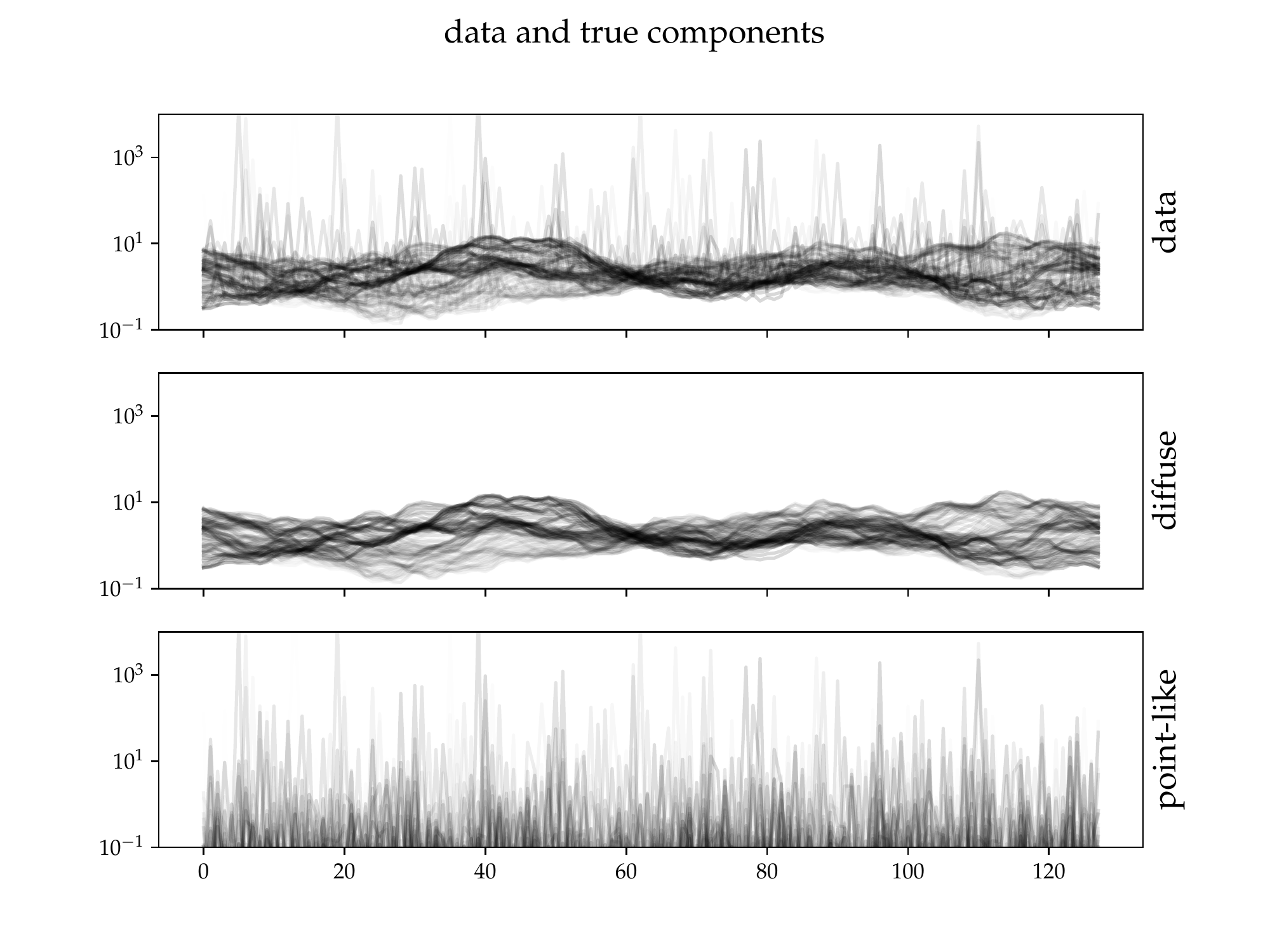}
	\caption{synthetic data (top), together with its diffuse component (middle) and point-like component (bottom) on a logarithmic scale, collapsed along one spatial dimension.}
	\label{fig:1d_data}
\end{figure}

For this comparison we generate the logarithmic diffuse component from a Gaussian process with the correlation structure
\begin{align}
p(k) = \frac{1}{(1+k)^{4}} \text{.}
\end{align}
The $k$ argument corresponds to the harmonic mode. It follows a power law with power $4$ which is equivalent to a smooth behavior in terms of small spatial curvature.
The point sources are drawn from the inverse gamma distribution with shape $\alpha = 1.5$ and scale $q =10^{-3}$. Both components are added together to generate our mock data.
The data can be seen in Fig. \ref{fig:2d_data}. In order to enhance the perception of the point sources, as well as the diffuse background we will look at the components edge-on. In order to do this we collapse the image along one direction and look at the brightness orthogonal to the collapsed direction. To obtain the visual effect of depth we increase the transparency linearly towards more distant locations, therefore faint features belong to the most distant locations along the collapsed direction, while saturated lines are close by. Our data, as well as the true diffuse and point-like component can be seen in Fig. \ref{fig:1d_data} in this representation. 
We apply the three different configurations of the \texttt{starblade} algorithm to the data. These three scenarios differ in the choice of $\alpha$. 

Its value in the first case is $\alpha=1.0$, which corresponds to an uninformative shape parameter. The prior distribution strongly favors bright sources, so it is easy for the algorithm to explain features with point sources. In the result we therefore expect diffuse contributions within the point-like component and a overly-smoothed diffuse component with underestimated power on small scales. In order to justify the variational approach we will also solve this configuration for its MAP solution and compare the results.

In the second scenario we pick the correct value for $\alpha=1.5$ and we therefore expect an excellent separation between the two components.

In the last configuration we choose a value of $\alpha = 3.0$, which strongly suppresses point sources, so the balance should lean towards more flux in the diffuse component, which will pick up point-source contributions. In this case we expect more power on small scales of the diffuse flux and a lack of faint point sources. It will still be easy for the algorithm to identify bright point sources, as they are absolutely incompatible with the diffuse flux. 

To compare the performance of our algorithm with other methods we first chose the background estimation step of the SExtractor \citep{sextractor}. SExtractor is a tool to extract sources from images and turn them into catalogs. In order to achieve this it performs a number of consecutive step, one being the subtraction of the image background, which corresponds to diffuse emission present in the image. This is done via $\kappa$-$\sigma$-clipping, which is iteratively performed on patches in the image. Within each patch a constant local background is determined, to which a median filter is applied to obtain a smooth background estimate of the whole image. Crucial to the outcome of this procedure is the choice of the patch size. The smaller it is, the more structures it can pick up. This behavior might not always be desired, as sources could be absorbed in the background. On the contrary, being too restrictive to a varying background, some of its features are identified as sources. The background window size by default is $64\times 64$ pixel. This will be our first SExtractor scenario. In order to tune it towards this problem, we will also reduce it to $8\times 8$ pixel. This is significantly smaller than the recommended range of $32$ to $128$ \citep{sextractor}.

Finally we train a denoising convolutional auto-encoder (DAE) on exactly this model. This kind of neural network specializes in removing noise or artifacts from images. It is trained on artificially corrupted images and its ground truth, in our case it gets the data and has to recover the diffuse component. Note that by its training with mock data from the correct model, the DAE was informed about the correct point source brightness distribution as well as about the correlation structure of the diffuse component. A detailed description of the network architecture and training is provided in Appendix \ref{ap:DAE}.

\begin{figure}
	\includegraphics[scale=0.45]{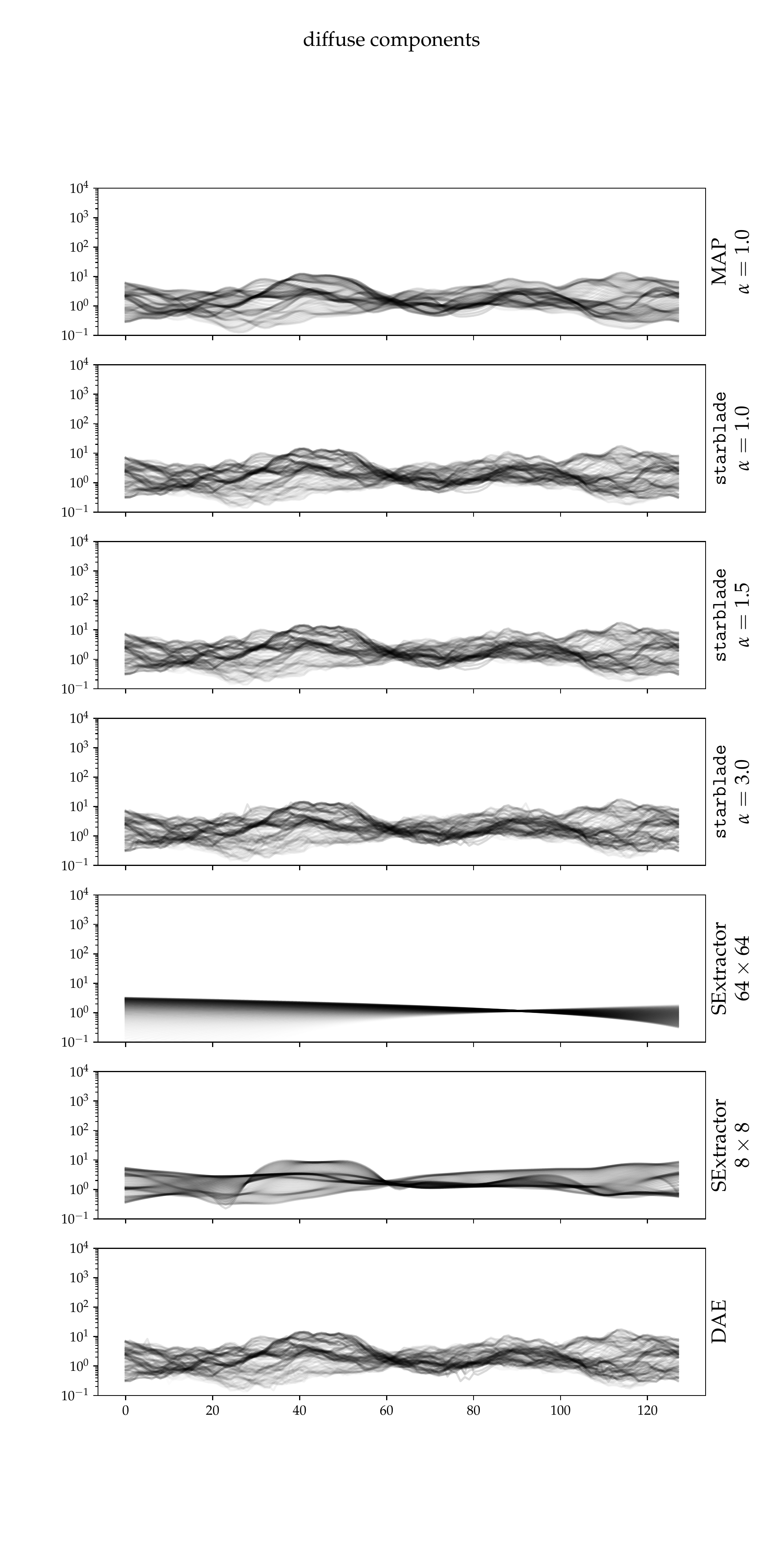}
	\caption{Results for the recovered diffuse emission for \texttt{starblade} with $\alpha = 1.0$, $1.5$ or $3.0$, respectively, on logarithmic scale, as well as for the two configurations of SExtractor, the DAE and the MAP solution for $\alpha=1.0$.}
	\label{fig:1d_diffuse}
\end{figure}

\begin{figure}
	\includegraphics[scale=0.45]{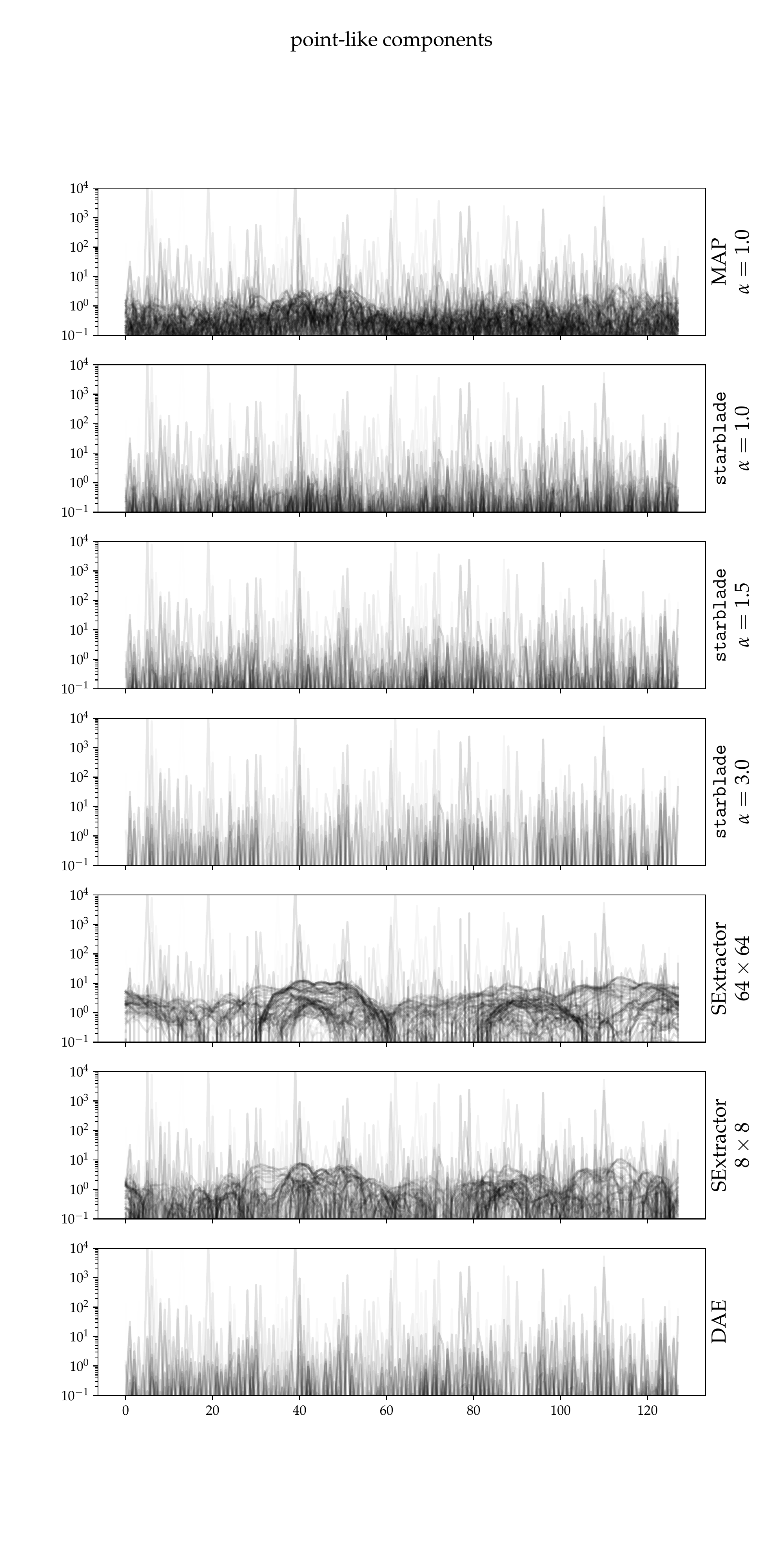}
	\caption{The recovered point-like flux for the set of algorithms on logarithmic scale.}
	\label{fig:1d_points}
\end{figure}

The results of the component separations for all these methods and configurations can be seen in Fig. \ref{fig:1d_diffuse} and Fig. \ref{fig:1d_points}, which show the resulting diffuse component and the point-like component respectively.

\begin{figure}
	\includegraphics[scale=0.6]{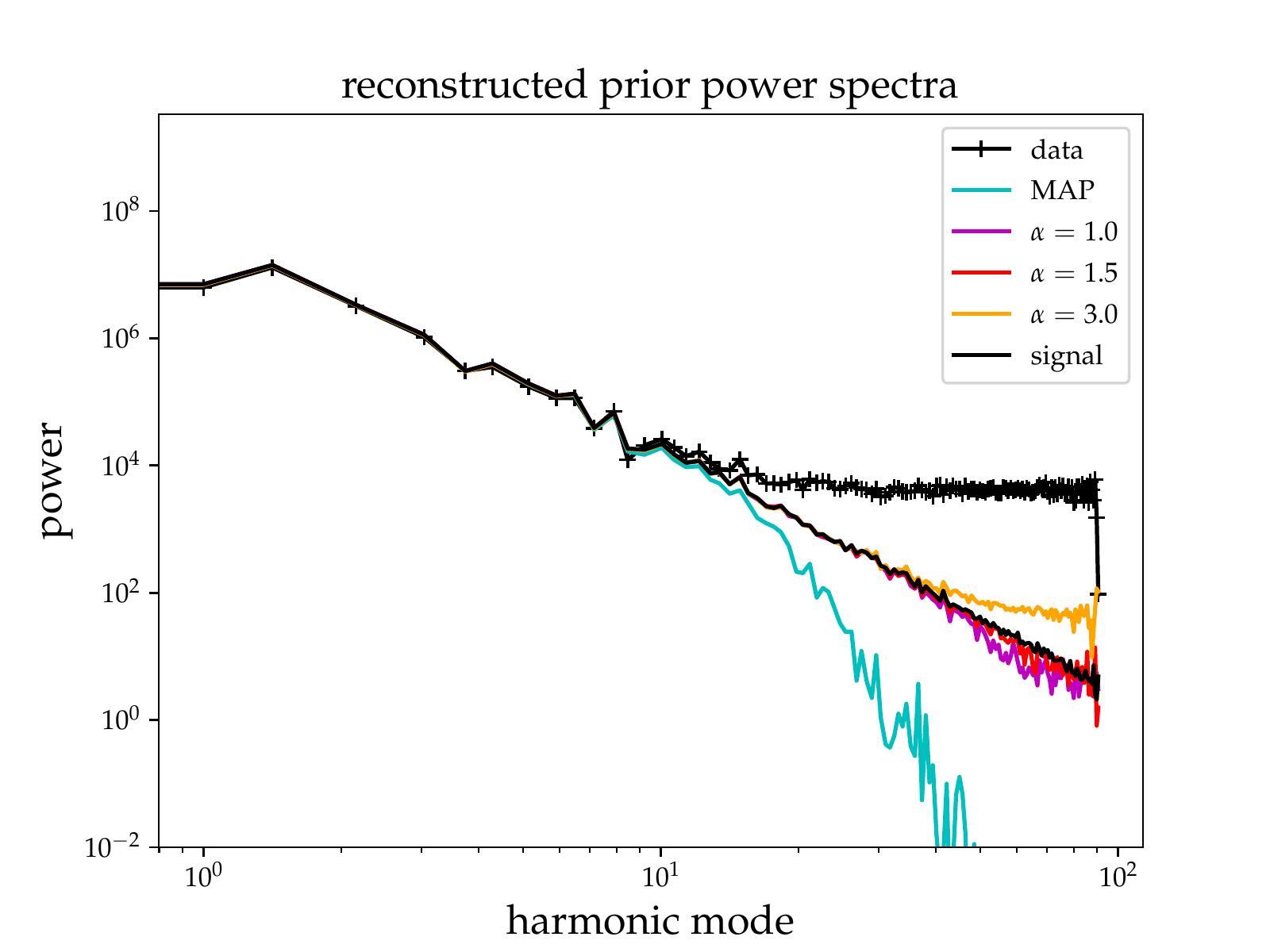}
	\caption{The recovered power spectra of the logarithmic diffuse component for the three different choices for $\alpha$ on double logarithmic scale, together with the power spectrum to create the diffuse component and the power of the logarithmic data.}
	\label{fig:1d_power}
\end{figure}

For the choice for $\alpha = 1.0$ in the \texttt{starblade} algorithm we obtain a diffuse component with correct large scale features, but it slightly lacks smaller scales and is slightly smoother than the original component. All these small scale features can be found in the point sources. Here a denser forest of small scales are visible. The brighter point sources are recovered correctly. In Fig. \ref{fig:1d_power} we see the results for the also reconstructed power spectrum, which characterizes the correlation structure of the underlying Gaussian process. In the first case  for $\alpha=1.0$, small scales are also slightly stronger suppressed, while the larger scales are recovered correctly. Comparing this result to the MAP solution, strong deviations become apparent. The recovered diffuse emission in this case is visually smoother and the point sources pick up a significant amount of small scale diffuse flux. The reconstructed power spectrum for the MAP solution drops off strongly towards the small scales as well. The minimization of the KL provides therefore a more robust result against an inappropriately chosen hyper parameter, compared to the minimization of the Hamiltonian.

The case with the correct $\alpha=1.5$ shows an excellent separation. Comparing the result with the true components, we do not find much difference. Neither remain obvious point sources in the diffuse component, nor the other way around. The recovered power spectrum of the diffuse component is spot on the correct one as well. This result verifies the correctness of our implementation of the \texttt{starblade} algorithm. We should mention here, that the MAP solution for the correct $\alpha=1.5$ gives only slightly worse results in this situation. We would expect a stronger difference in situations with more point-source flux, which corresponds to a higher noise on the diffuse emission.

\begin{figure}
\includegraphics[scale=0.6]{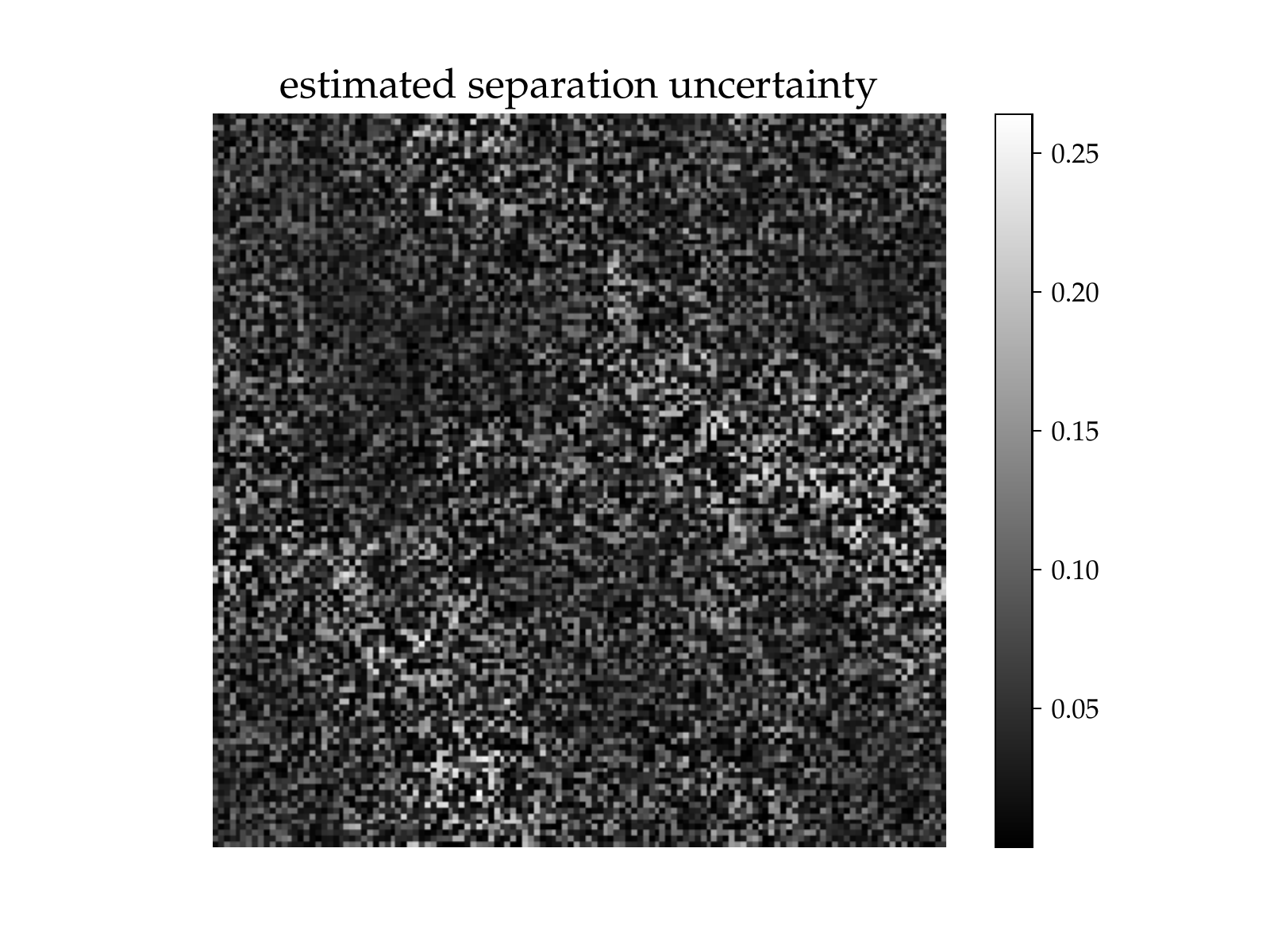}
\caption{Estimated uncertainty in terms of standard deviation of the separation for $\alpha=1.5$.}
\label{fig:uncertainty}
\end{figure}

Our algorithm is additionally capable of providing uncertainty on its estimates. In Fig. \ref{fig:uncertainty} the uncertainty of the separation field $a$ is shown, which translates to an uncertainty on either component. Is shows large and small-scale features. The uncertainty is high in regions where both components appear strongly mixed. This can be seen by the large scales, which follow the diffuse component. In regions this component is weak, the uncertainty drops down and the algorithm is confident of its separation. 

The final \texttt{starblade} scenario with $\alpha = 3.0$ also shows a reasonable separation. As expected the the diffuse component exhibits more small scale features compared to lower $\alpha$ and the point sources appear thinned out at the faint end. This also reflects within the reconstructed power spectrum. For small scales it exhibits higher power compared to the true underlying signal as the missed faint point sources are absorbed in the diffuse component.

Applying the SExtractor background estimation with a patch size of $64\times64$ pixel does not provide a reasonable component separation, as a significant portion of the diffuse emission remains within the separated point sources. The default settings of SExtractor are not reasonably applicable in this situation, as it uses a patch size of $64\times 64$ pixel, which corresponds to four patches over the test image, so one cannot expect a detailed separation. The choice of $8\times 8$ patches performs significantly better. It is capable to at least resolve large scale features within the diffuse emission, but still attributes smaller scale correlated features to the point-like emission. 

Finally, the last method we want to compare our method to is the specially trained DAE. It is worth to note that during the training the correct correlation structure was used, compared to the \texttt{starblade} method, which was agnostic to it. The auto-encoder therefore was equipped with an advantage concerning the a priori knowledge on the problem. The method performs excellent as well. The results, at least by eye, are comparable to the ones obtained by our method. 

\begin{figure}
	\includegraphics[scale=0.6]{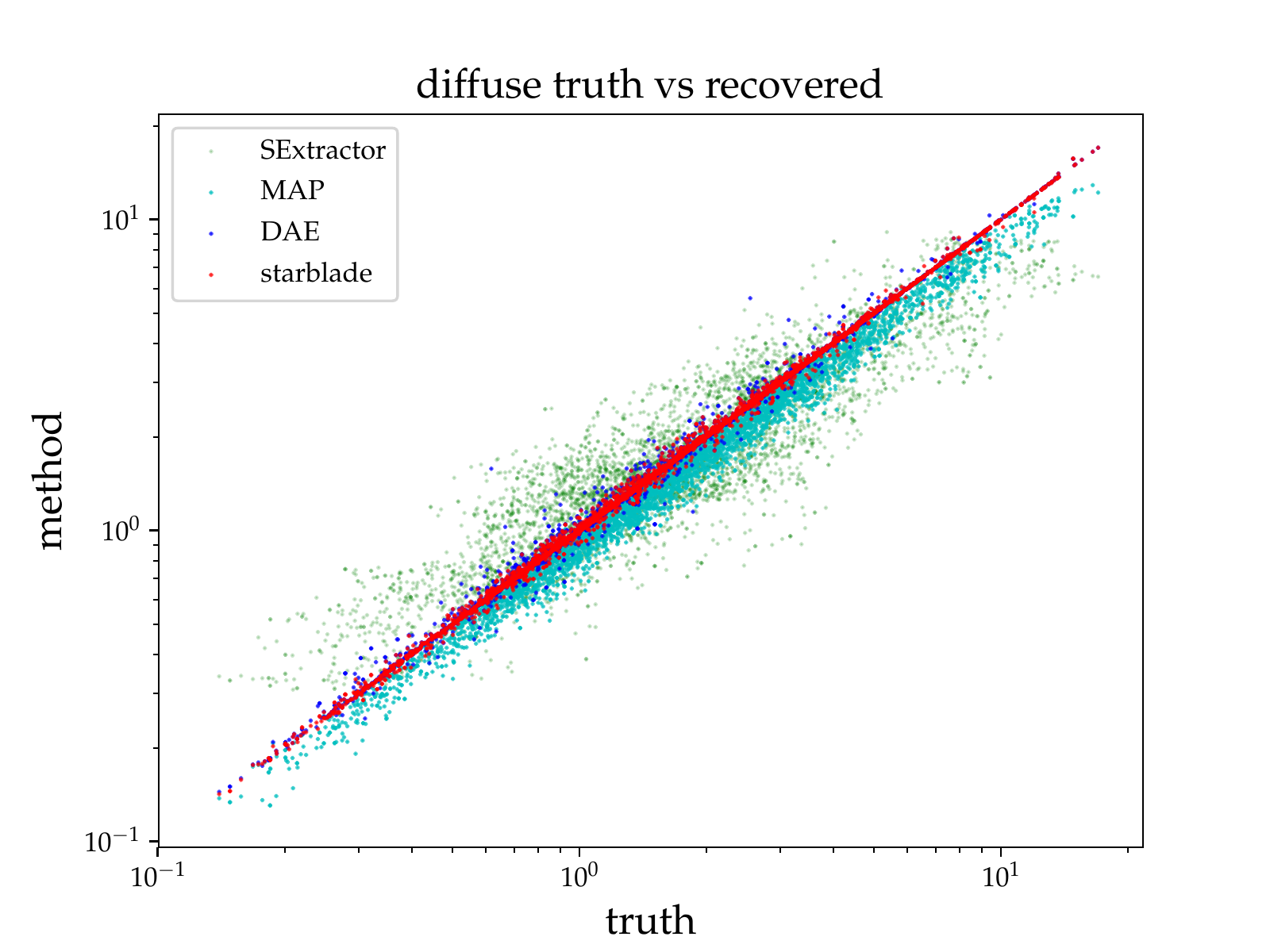}
	\caption{The true diffuse component plotted against the separated for \texttt{starblade} with $\alpha=1.5$, DAE, MAP with $\alpha=1$ and SExtractor using $8\times 8$ pixel patches.}
	\label{fig:scatter}
\end{figure}

To further  investigate the difference between the methods we plotted the results for the diffuse components pixel-wise versus the true underlying component, which is shown in \ref{fig:scatter}. For this plot we sampled a subset of random locations and for a perfect separation we expect a diagonal line. Here we only used the best performing versions of each method, namely \texttt{starblade} with $\alpha = 1.5$ and SExtractor with $8\times 8$ and the auto-encoder. Additionally we also plotted the MAP solution with $\alpha=1.0$ to show the sensitivity of MAP to an suboptimal hyper parameter. We see that SExtractor scatters the most and it tends to completely cut low and high flux diffuse emission.  The DAE performs a lot better with significantly lower scattering. It is also capable to identify high-flux diffuse areas. Here the \texttt{starblade} method exhibits an even lower variance. It builds almost a straight line. The MAP result is systematically shifted towards lower flux in the diffuse component, which reflects the higher assumed point-source flux, expressed in the lower $\alpha$. To quantify all those differences we calculate the root mean squared error (RMS) on this logarithmic scale of the deviations of the result compared to the truth. The RMS values for all methods are displayed in Table \ref{tab:example_table}.

\begin{table}
	\centering
	\caption{The RMS error of the logarithmic classification of all methods and configurations.}
	\label{tab:example_table}
	\begin{tabular}{|l|l|} 
		\hline
		Method & RMS Error\\
		\hline
		MAP $\alpha=1.0$ & $0.15$\\
		\texttt{starblade} $\alpha=1.0$ & $0.035$\\
		\texttt{starblade} $\alpha=1.5$ & $0.026$\\
		\texttt{starblade} $\alpha=3.0$ & $0.056$\\
		SExtractor $64\times64$ & $1.4$\\
		SExtractor $8\times8$ & $0.35$\\
		 DAE & $0.049$\\
		\hline
	\end{tabular}
\end{table}

The RMS error is the highest for both SExtractor configurations, which have an error one or two orders of magnitude higher compared to the other methods. The sub-optimal choices for $\alpha$ in the \texttt{starblade} algorithm perform similar to the to the DAE, where $\alpha=3.0$ is slightly worse and $\alpha=1.0$ slightly better. The least error is accomplished by the \texttt{starblade} algorithm with the optimal choice for $\alpha = 1.5$. It achieves half the error compared to the specially trained network. In this task our method is superior compared to any other tested methods. Other network architectures might achieve a better result, but increasing the accuracy by another factor of two would probably require serious effort. 

Overall each of the presented methods has its own advantages and disadvantages. The SExtractor background estimation is extremely fast and robust, but lacks precision. It might be sufficient for a large number of applications, but if higher accuracy is required one might want to use another background estimation. The DAE performs reasonably well and is easy to implement and set up. It performs within the same magnitude as \texttt{starblade}. Training the network requires some time, but after that the separation is done quickly. The reasoning of the network for its conclusion is, however, nebulous. More sophisticated architectures might have an increased performance, but they do not origin from first principles and can only be obtained via experimentation.
In contrast to that, \texttt{starblade} is derived from probability theoretical considerations. The model assumptions are physically motivated. Compared to the previously mentioned methods we can also provide an estimate of uncertainty for the separation. For this method no training phase is required, but the separation procedure itself requires higher computational effort. In some cases one might be able to take the shortcut of the MAP solution, which can be calculated significantly faster, but this requires a careful selection of the $\alpha$ parameter.

\subsection{$\mathrm{M}100$ observed by the Hubble Space Telescope}
\label{sec:hubble}

So far we only considered synthetic data, for which the ground truth is known. There we could compare the different methods directly to the truth. This is no longer possible for real data applications. Here we can only describe the differences within the methods and judge the performance subjectively. In case of real data, the idealistic model assumptions do not hold any more. We will investigate  how well the method generalizes. To do this we apply all previously used methods to separate an image obtained by the Wide-Field Planetary Camera 2 (WFPC2) mounted to the Hubble Space Telescope of the galaxy $\mathrm{M}100$ \citep{M100}.

\begin{figure}
	\includegraphics[scale=0.6]{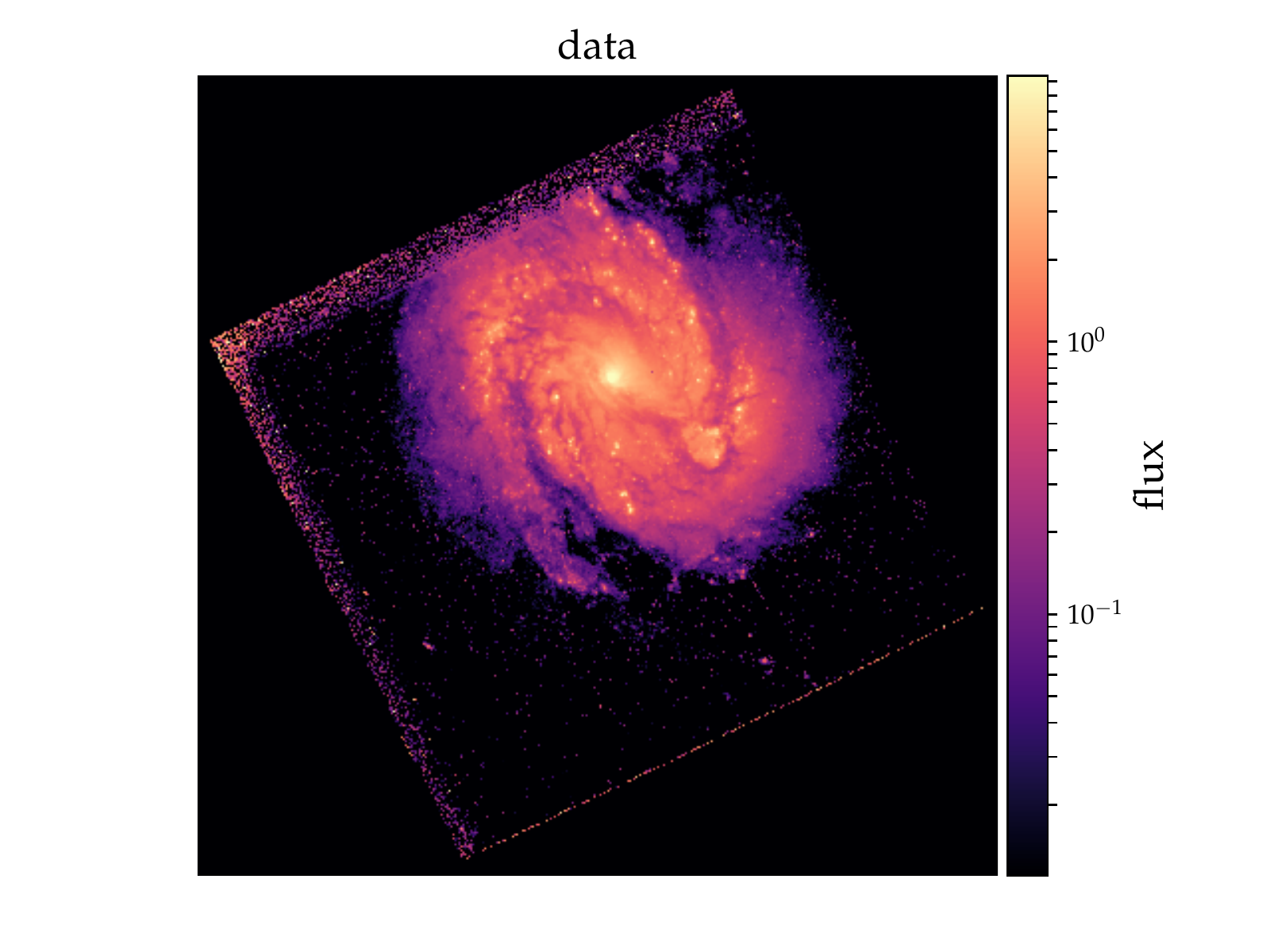}
	\caption{The data of the $\mathrm{M}100$ galaxy on logarithmic scale.}
	\label{fig:M100_data}
\end{figure}

The image is subject to noise, convolved with a point spread function, affected by cosmic ray hits and exhibits regions with high noise levels at the edge of the field of view. The logarithmic data is shown in Fig. \ref{fig:M100_data}. Because of the point-spread function, bright sources are smeared out over a larger area, which reduces the brightness within individual pixels, the canonical choice for $\alpha = 1.5$ is too restrictive for these spread point-sources.  They  tend to be absorbed within the diffuse component. To counteract this behavior we reduce its value to $\alpha = 1.1$. Everything not compatible with diffuse flux will be part of the point-like component, therefore we wish to separate the diffuse component efficiently from foreground stars, cosmic ray hits and noise artifacts.

\begin{figure}
	\includegraphics[scale=0.6]{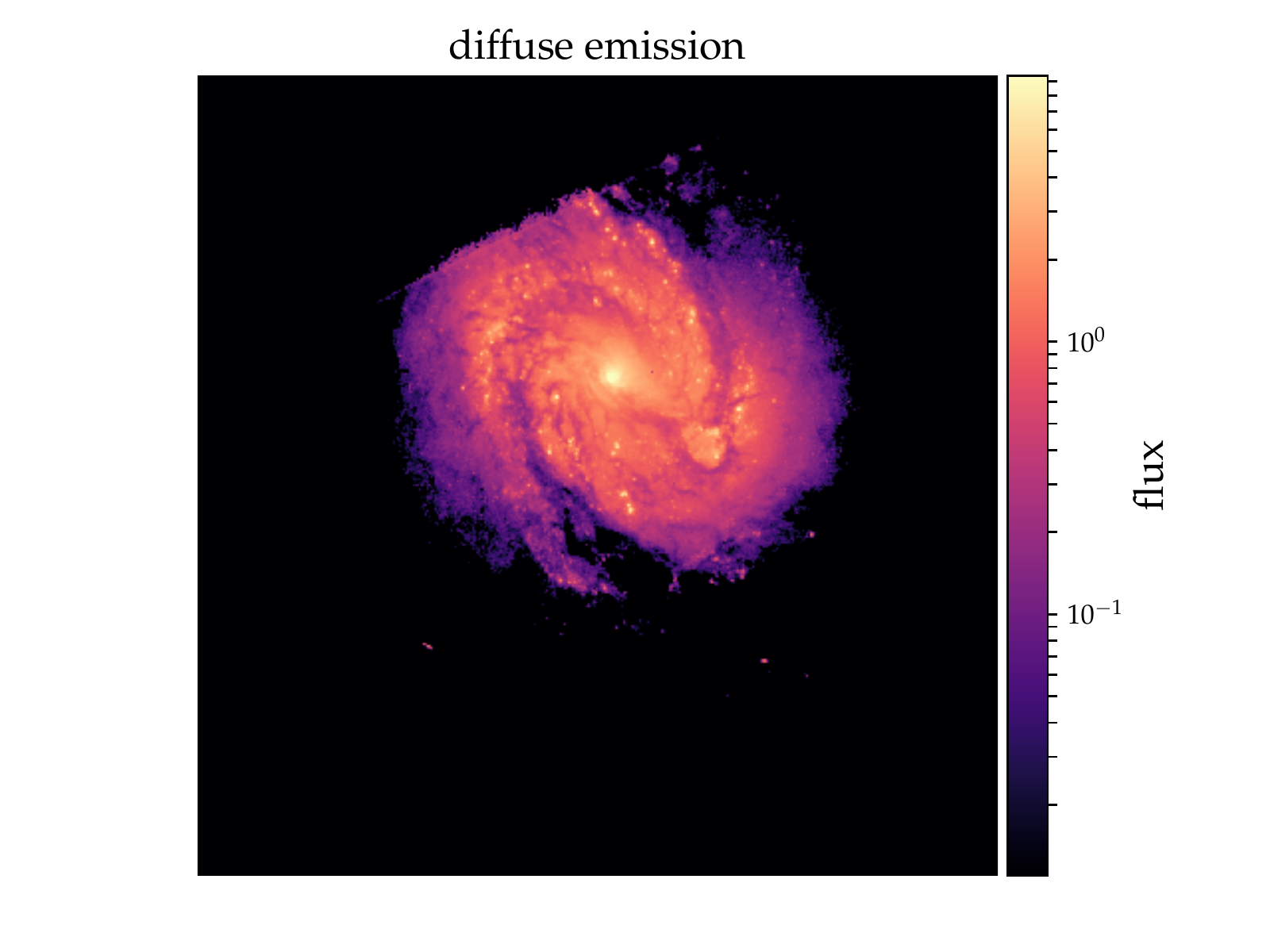} 
	\caption{The separated diffuse component on logarithmic scale.}
	
	\label{fig:M100_diffuse}
	
\end{figure}

\begin{figure}
	\includegraphics[scale=0.6]{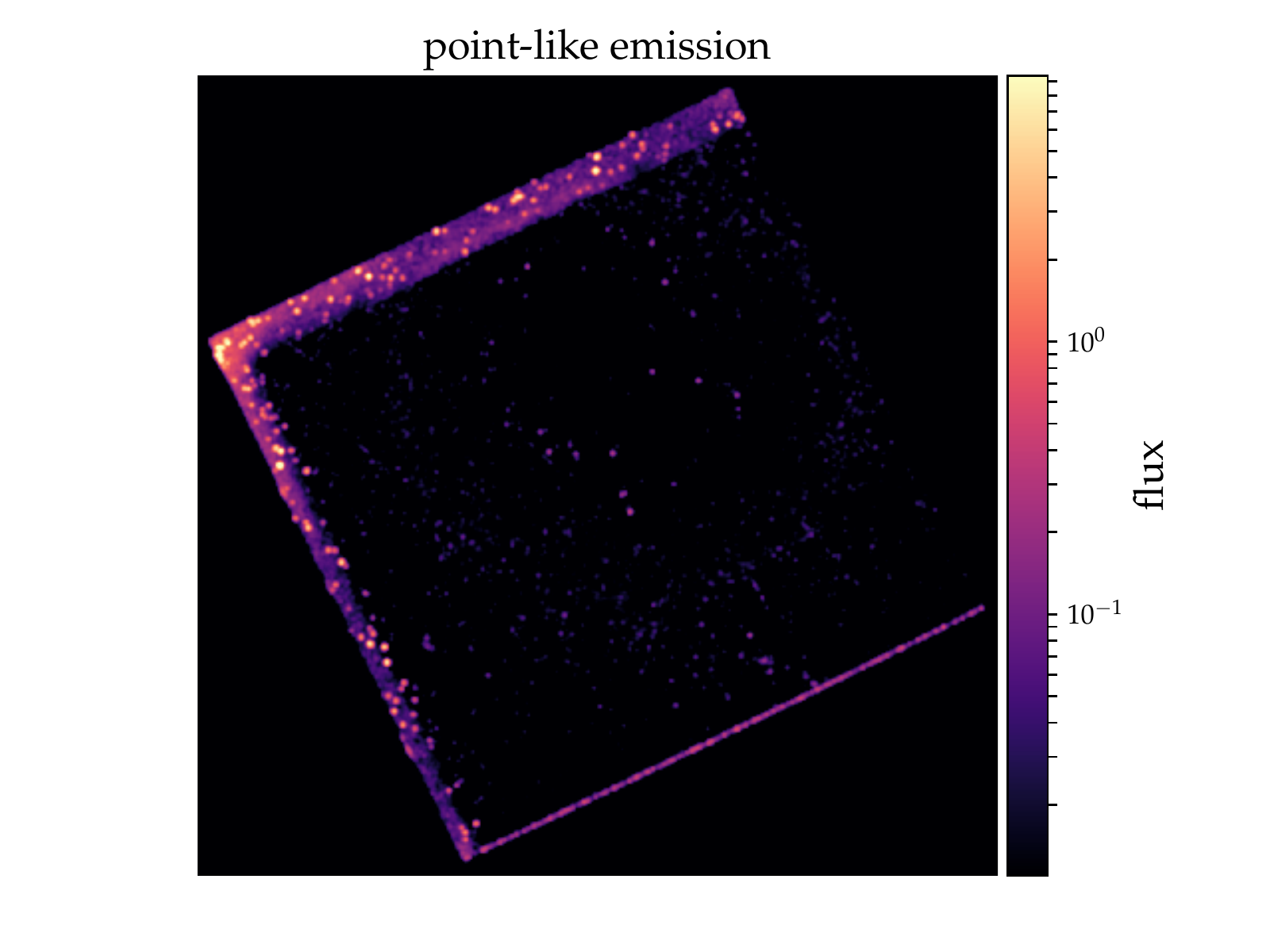}
	\caption{The separated point-like component on logarithmic scale. The linear flux image has been convolved with a Gaussian beam to enhance the visibility of the separated point sources.}
	
	\label{fig:M100_point}

\end{figure}

\begin{figure}
	\includegraphics[scale=0.6]{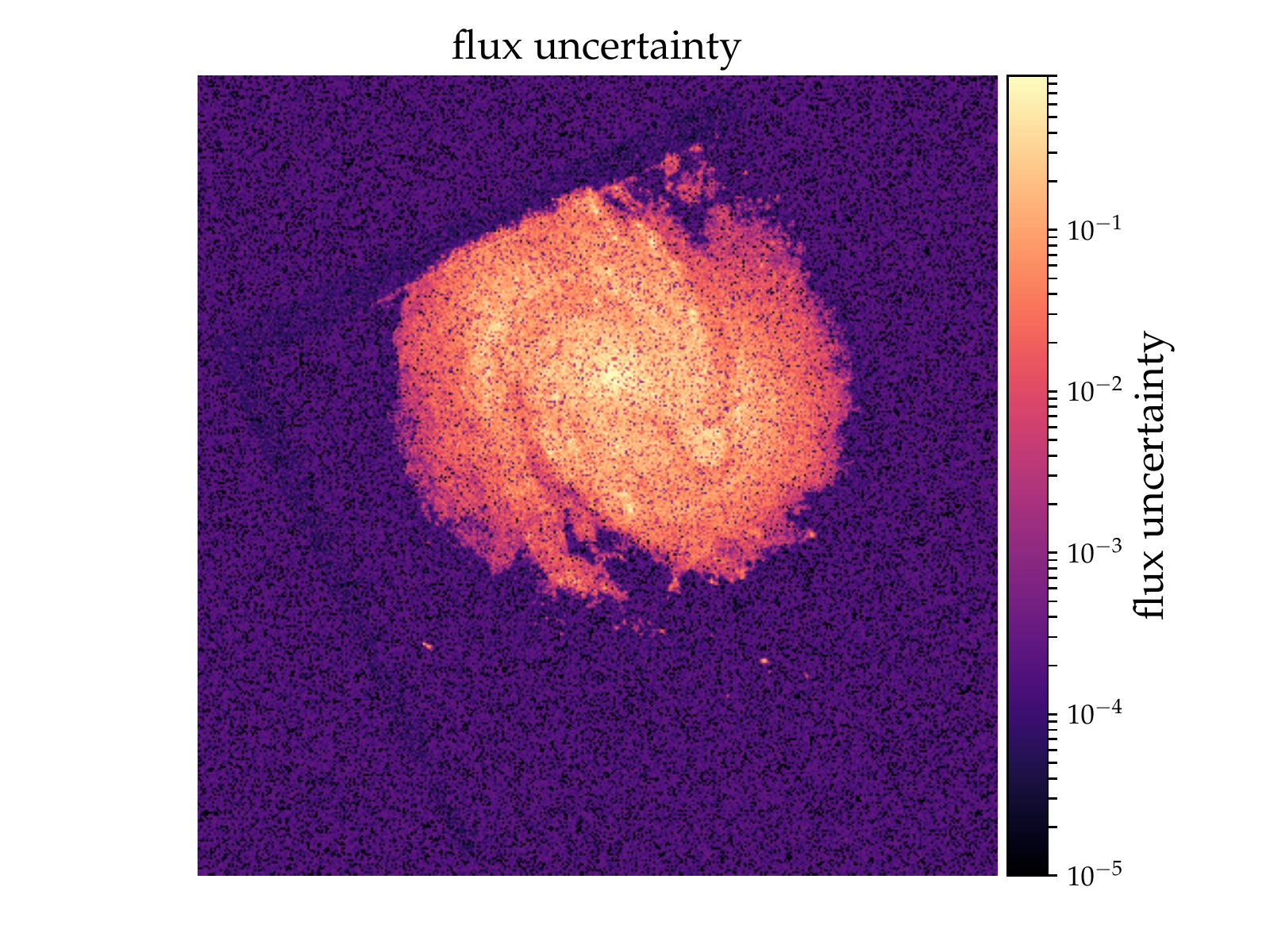} 
	\caption{The flux uncertainty in terms of a one sigma interval.}
	
	\label{fig:M100_uncertainty}
	
\end{figure}

\begin{figure}
	\includegraphics[scale=0.6]{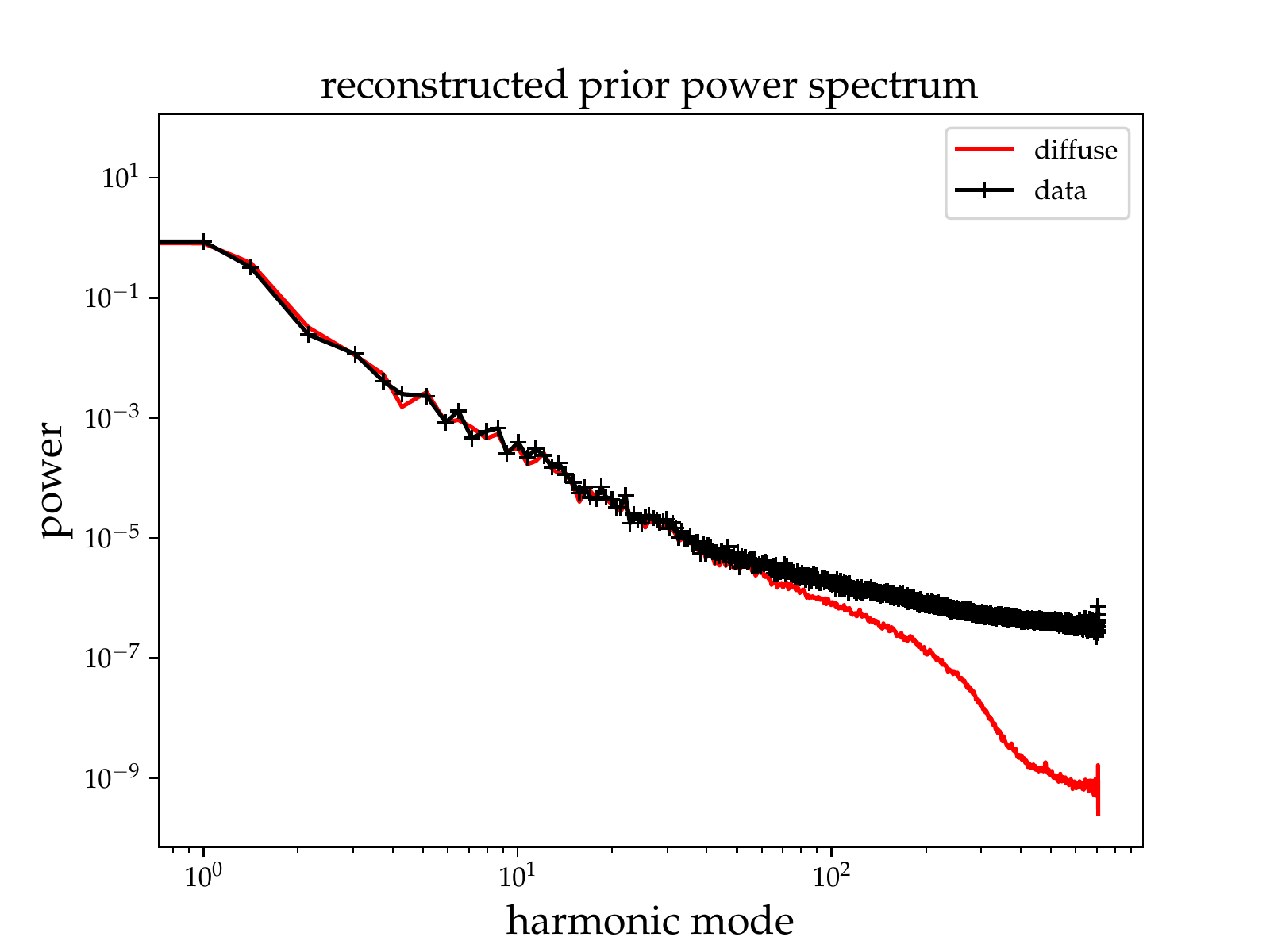}
	\caption{The recovered power spectrum of the logarithmic diffuse component with the power spectrum of the logarithmic data.}
	
	\label{fig:M100_power}
	
\end{figure}

The recovered diffuse component can be seen in Fig. \ref{fig:M100_diffuse}. Almost any flux outside the disk of $\mathrm{M}100$ itself was identified as point-like emission and removed from the diffuse component. The brightest points inside the disk were removed as well. What remains is the diffuse emission from the galaxy. The recovered point sources are shown in Fig. \ref{fig:M100_point}. This image was convolved with a small Gaussian kernel to enhance the visibility of the point-sources. Here we clearly obtain the brightest sources, some of them superimposed on the diffuse structure. Additionally,   most measurement artifacts are captured by the point source component, such as the rectangular edge of the field of view, as they are incompatible with diffuse emission. The estimated flux uncertainty can be seen in Fig. \ref{fig:M100_uncertainty}, which shows the expected standard deviation associated with every location. It clearly follows the diffuse component. This is reasonable, as we do expect higher uncertainties at locations where both components are superimposed.  The recovered power spectrum of the logarithmic diffuse emission can be seen in Fig. \ref{fig:M100_power}, as well as the power of the logarithmic data. The data on large scales are dominated by the diffuse emission, the small scales by the point sources. The correlation structure of the diffuse emission obtains all the power on large  and intermediate scales and continues to drop with a constant slope towards small scales, in contrast to the data.

\begin{figure}
	\includegraphics[scale=0.6]{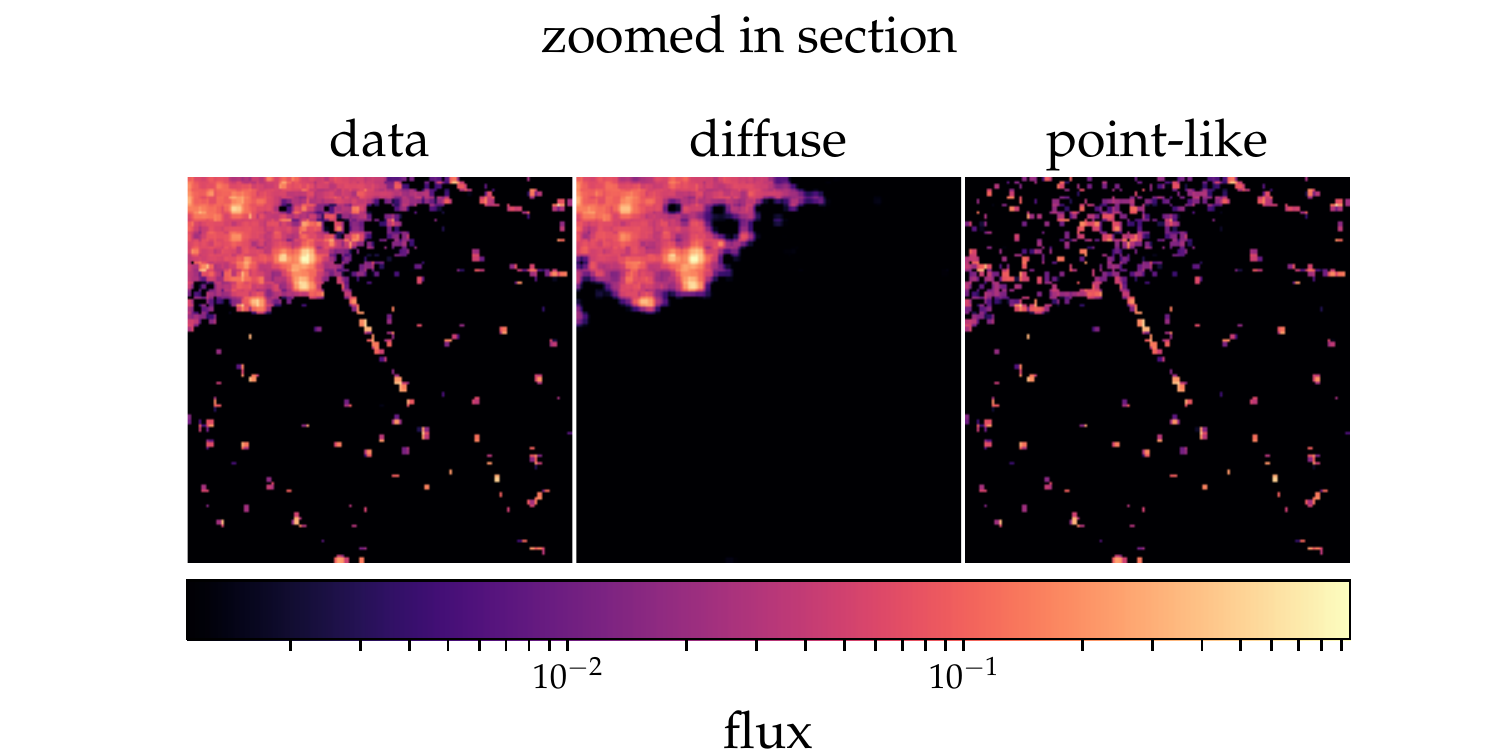}
	\caption{A zoomed in section for the data and the separated components on logarithmic scale. Here the point sources are not convolved.}
	
	\label{fig:M100_zoom}
	
\end{figure}

We can also detect cosmic ray hits in the form of point sources in a consecutive line in this component. One such example can be seen in Fig. \ref{fig:M100_zoom}, which shows a zoomed in section of the image on the edge of the disk of $\mathrm{M}100$. Here the recovered point sources are not convolved artificially. Even though a point spread function of the instrument is present, and point sources do not only inhabit individual pixels in the image, the difference between the detected diffuse emission and smeared out high intensities from point sources is sufficient to separate both components, at least for the brightest sources.

Overall we obtained a good estimate of the diffuse emission of $\mathrm{M}100$, removing any point-like contribution, originating either from point-sources or from systematics.

\begin{figure}
	\includegraphics[scale=0.6]{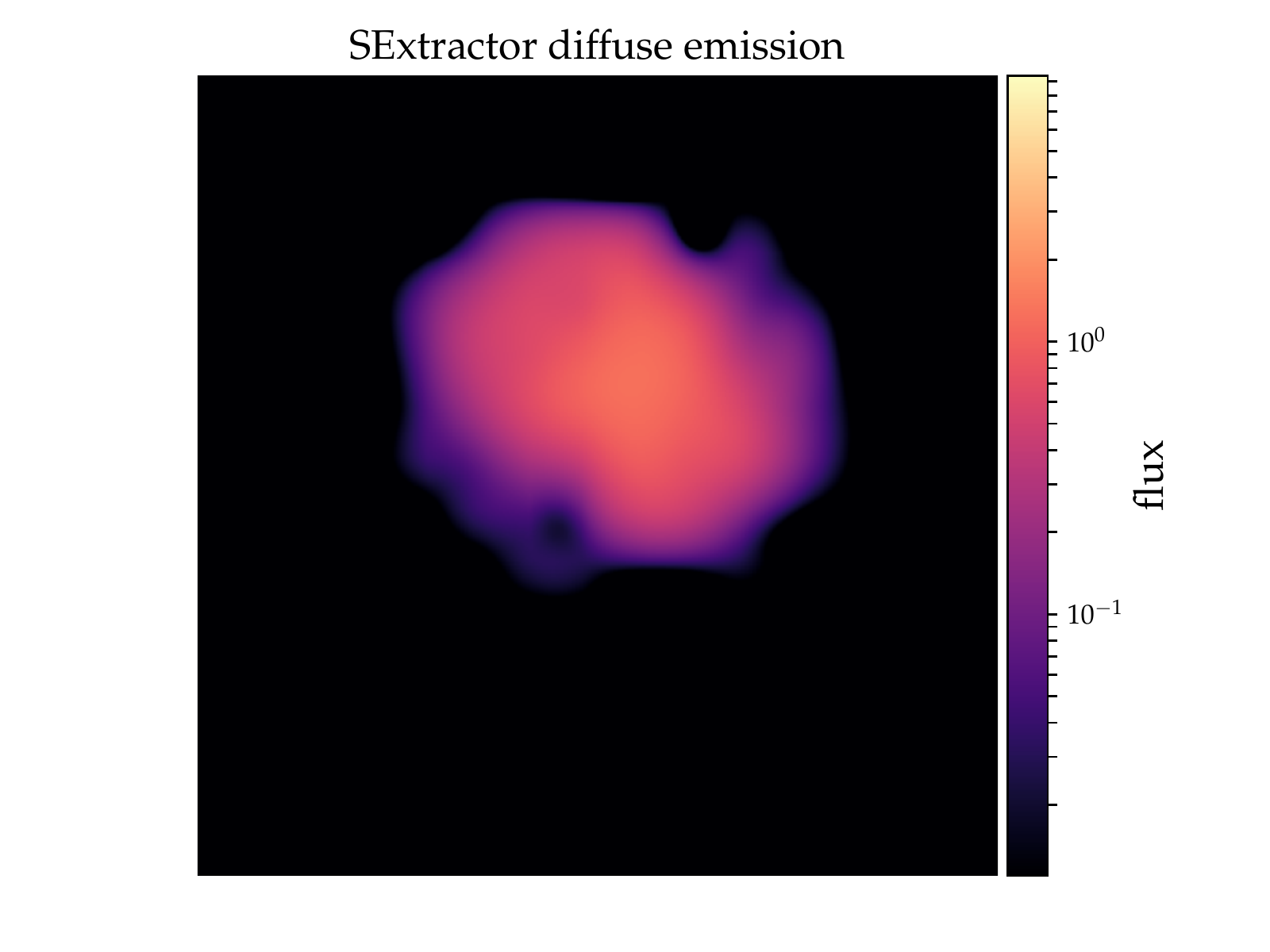}
	\caption{Diffuse component obtained by SExtractor with the $64\times 64$ pixel window.}
	\label{fig:SEhubblediff}
\end{figure}

\begin{figure}
	\includegraphics[scale=0.6]{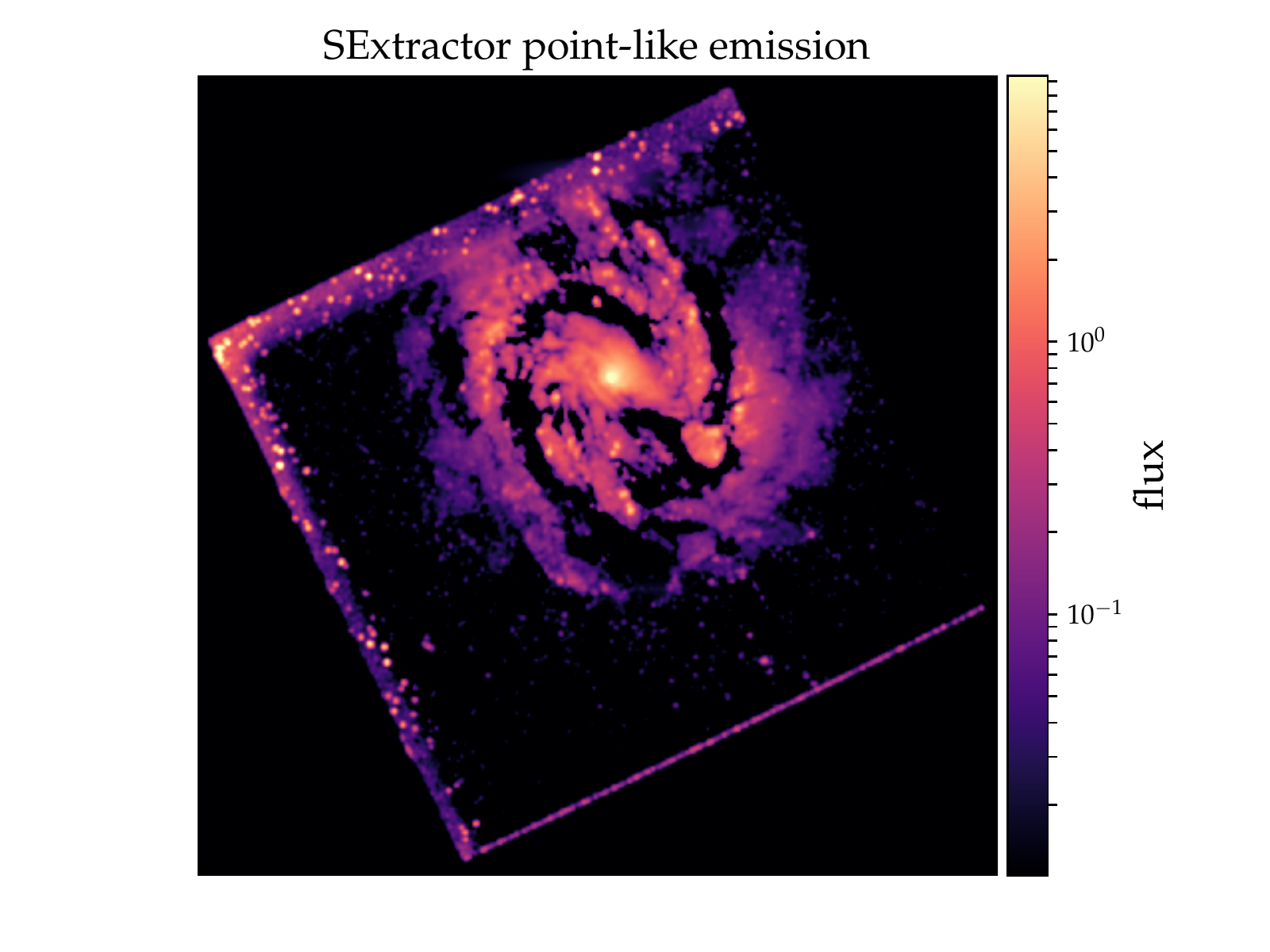}
	\caption{Convolved point-like component obtained by SExtractor with $\mathrm{BACKSIZE}=64\times 64$.}
	\label{fig:SEhubblepoint}
\end{figure}

The application of the background estimation of SExtractor does provide a less reasonable result. The largest structures are correctly identified as diffuse emission, as can be seen in Fig. \ref{fig:SEhubblediff}. Significant amounts of smaller structures, but still clearly diffuse emission, remains within the point-sources. This we already observed in the mock example. The point-like component can be seen in Fig. \ref{fig:SEhubblepoint}. The disk is split into several individual patches. Introducing such artifacts might be hard to deal with in further analysis. Applying the SExtractor background separation also introduces areas of negative flux in both components, which on the other side artificially creates flux. For illustration purposes this negative flux was clipped. An advantage of this approach is, that certainly no point-like flux remains within the diffuse component.

\begin{figure}
	\includegraphics[scale=0.6]{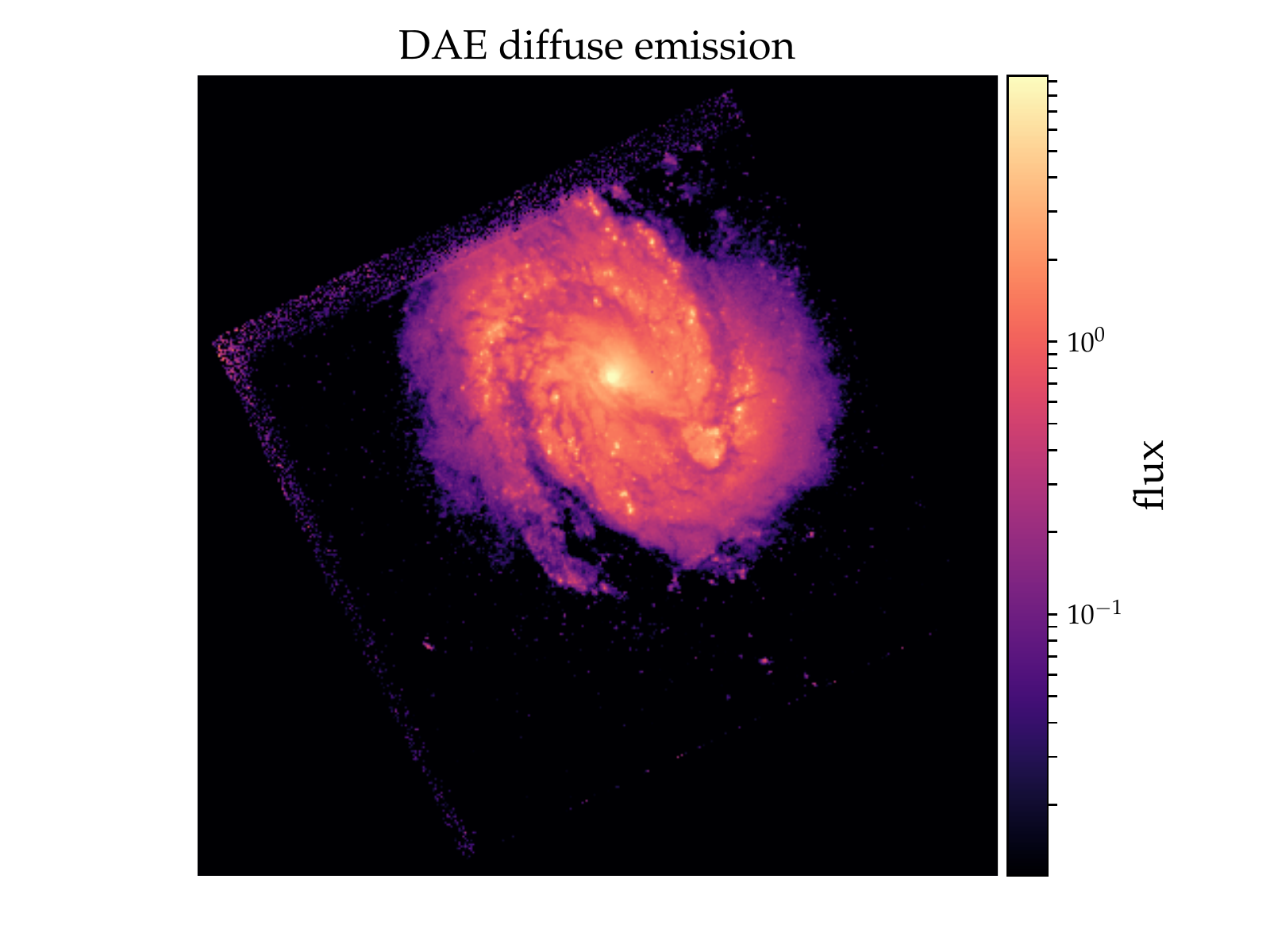}
	\caption{Diffuse component separated by the DAE.}
	\label{fig:DAEhubblediffuse}
\end{figure}

\begin{figure}
	\includegraphics[scale=0.6]{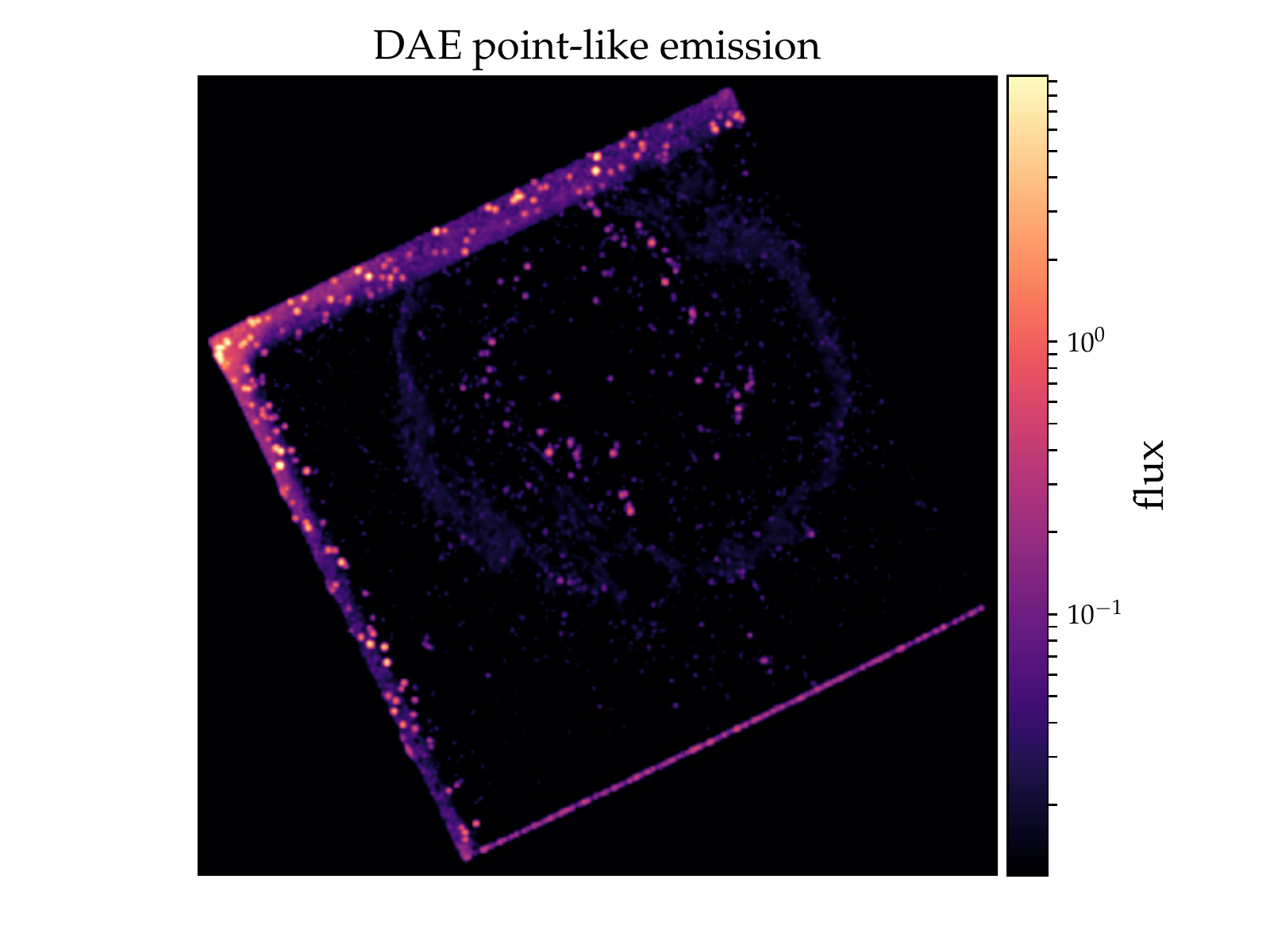}
	\caption{Convolved point-like component separated by the DAE.}
	\label{fig:DAEhubblepoint}
\end{figure}

Applying the identical DAE trained on the  previous model provides a relatively reasonable result, which can be seen in Fig. \ref{fig:DAEhubblediffuse} and Fig. \ref{fig:DAEhubblepoint}. The network therefore somehow abstracted the notion of point sources to a degree to make it applicable outside its training set. This observation is not trivial, as neural networks are not guaranteed at all to show this behavior. It performs well in subtracting the diffuse component from the point sources. It separates a larger amount of flux from the disk compared to \texttt{starblade}, which might still belong to the galaxy as they follow its morphology. The diffuse component, however still contains a significant amount of point-like emission. One cannot train a network directly on such data sets, as we do not have access to the  ground truth of the separation and in any case one has to rely on generating  a training set according to some model. Other network architectures also might lead to a better performance, but this requires a large amount of experimentation.

\begin{figure}
	\includegraphics[scale=0.6]{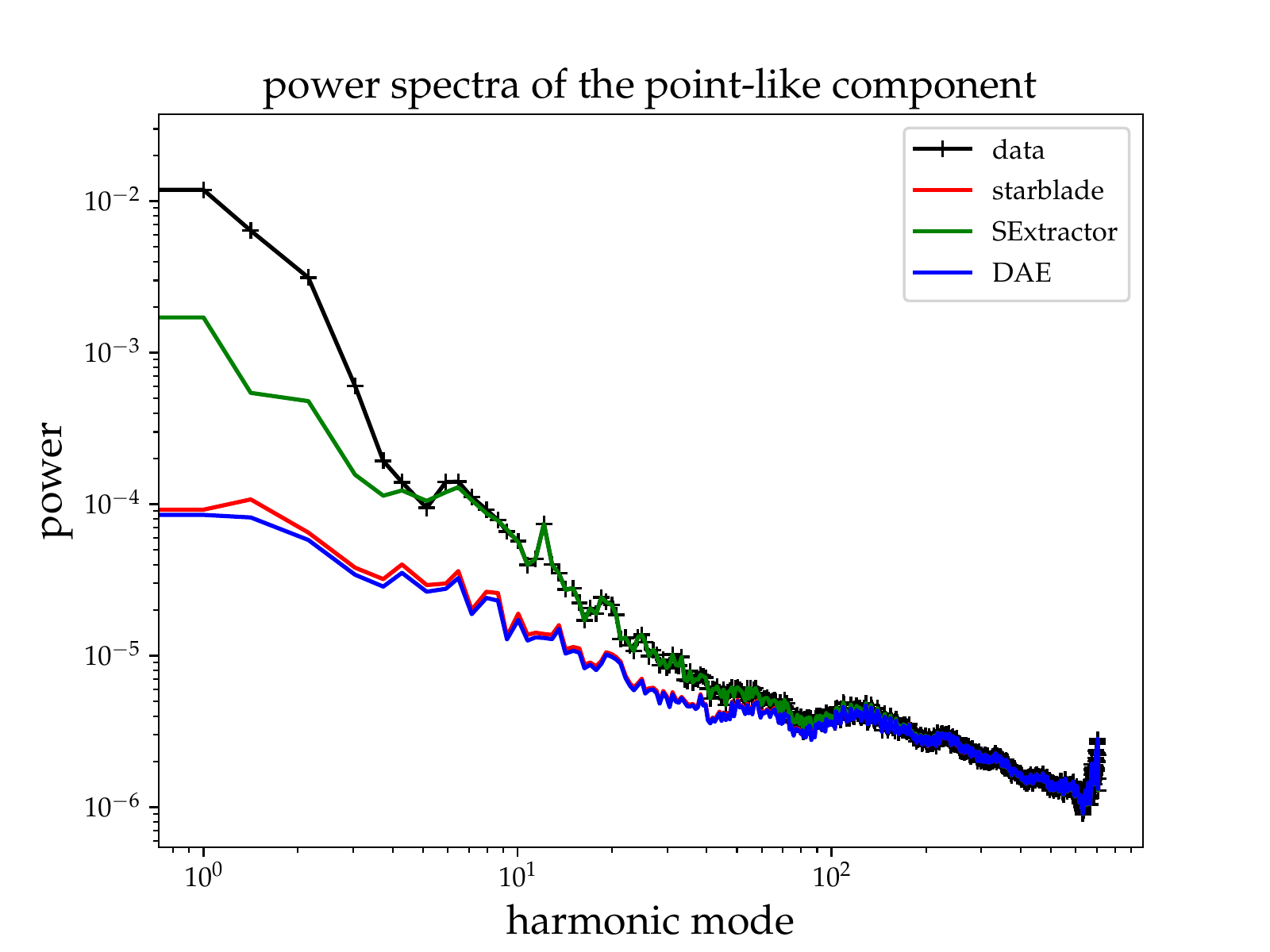}
	\caption{The power spectrum of the linear maps of the point-like components on double logarithmic scale.}
	\label{fig:pointpower}
\end{figure}

To compare the different results we can look at the power spectrum of the point-like components of the different methods. This time we calculate the power of the components on a linear scale. For randomly scattered point sources we would expect a flat power spectrum with equal power on all scales. Deviations from that indicate an incomplete separation. Especially high power on large scales correspond to a remaining large-scale background. The power spectra can be seen in Fig. \ref{fig:pointpower}. The data itself exhibits power on large scales, which drops of and then flattens out. The large scales are dominated by the spatial extension of the galaxy, while the flattening can be attributed to the point sources. The increase at the smallest scales show the point-spread function of the instrument itself. Subtracting the background obtained by SExtractor removes roughly one order of magnitude in power for the largest scales, but everything below some large threshold is captured in this component. Overall the power spectrum has a large slope, which corresponds to a spatially correlated structure in this point-like component. Compared to this, the other two methods have a significantly lower slope, and therefore less large-scale structure. At the largest scales these components exhibit roughly two orders of magnitude less power and the spectrum stays below the power of the data even for smaller scales.

\begin{figure}
	\includegraphics[scale=0.6]{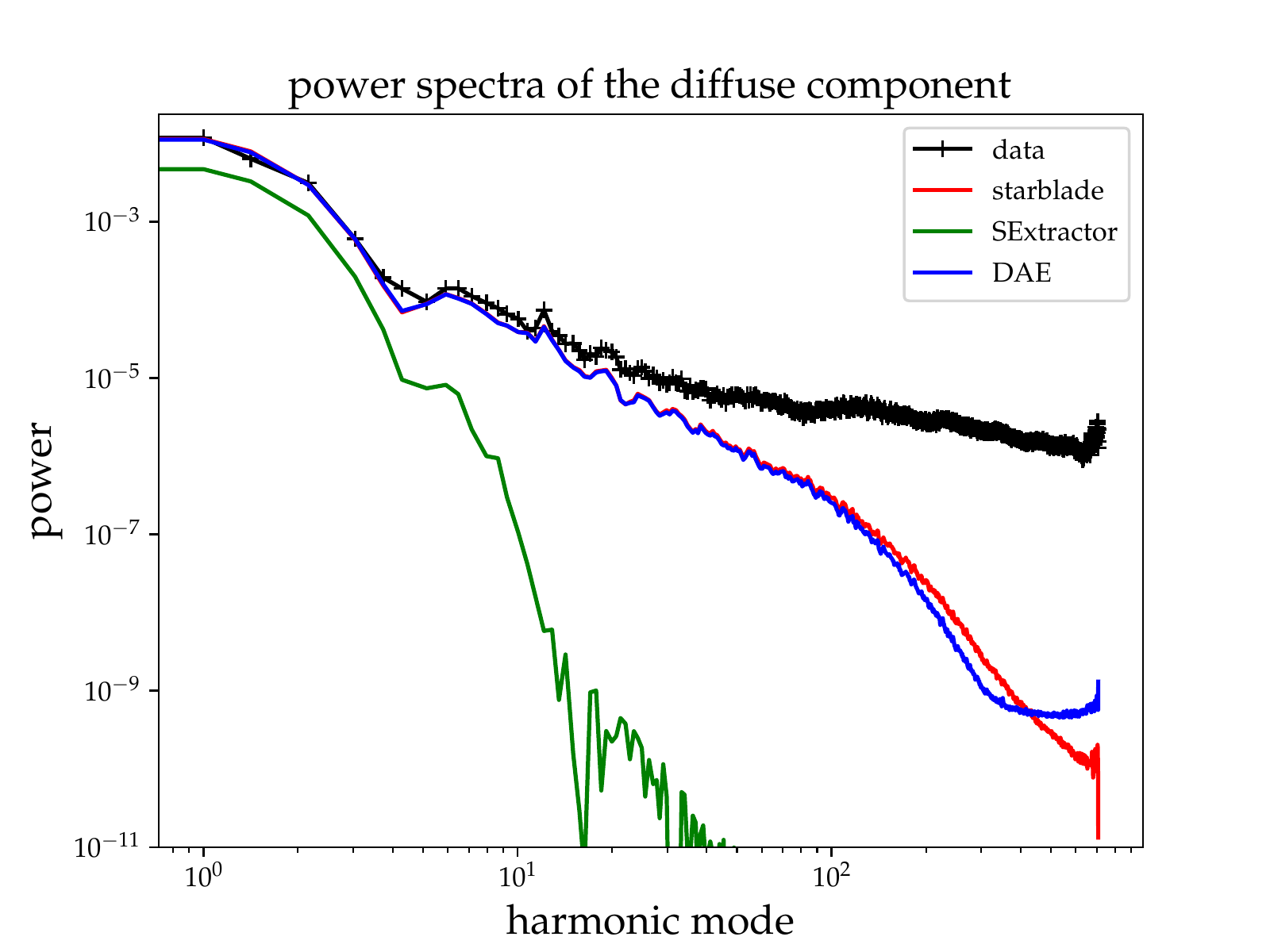}
	\caption{The power spectrum of the linear maps of the diffuse components on double logarithmic scale.}
	\label{fig:diffusepower}
\end{figure}

One can also look at the power spectrum of the recovered diffuse components. Here, falling power spectra indicate correlated structures. These are shown in Fig. \ref{fig:diffusepower}. The power spectrum of the background estimation of the SExtractor has large power on large scales, but is shifted down systematically compared to the data. Towards smaller scales it drops of rapidly, which does not allow for smaller scale correlated features. Compared to that, the other two methods explain large scales almost exclusively with the diffuse component. Once their power diverges from the data, the spectrum exhibits a series of bumps, which should correspond to some characteristic scales within the structure of the galaxy. In the data alone, these structures are hidden by the power of the point sources. At the smallest scales he DAE drops of slightly steeper and then levels off flat. This leveling off is a sign of remaining point sources of some smaller brightness, which we can also observe in the reconstructed images. The \texttt{starblade} algorithm does not show any leveling off, which indicates the absence of any point sources above the smallest scales of the diffuse component. At the very end of the spectrum, it deviates from the drop. This coincides with the increase in power of the data due to the point-spread function and we attribute it to this instrumental effects. Besides this, our algorithm seems to generalize well in this real data application.

To investigate further, we can look at a number of correlation metrics between the different components and methods. Initially we assumed the components to be independent of each other. One way to test this, is to calculate the correlation between the results of the separation. A vanishing correlation does not imply independence, but the reverse holds, so the more correlation we observe, the less independence in the separation we can assume. We will measure the correlation by their cosine similarity, which is given by
\begin{align}
\mathrm{cos}(\theta) = \frac{a^\dagger b}{\vert a\vert\vert b \vert} \text{.}
\end{align}
Here $a$ and $b$ are the components and $\theta$ defines the angle between them. Uncorrelated components are orthogonal to each other, and therefore the cosine similarity vanishes. Highly correlated components have a large overlap and therefore a small angle between them, leading to a similarity of unity. The cosine similarities between the separated components for the different methods can be found in Tab. \ref{tab:correlation_table}. The largest similarity is found in SExtractor. This is not too surprising, as both components exhibit a significant portion of large scale structure. One order of magnitude less similarity can be found in the DAE and a bit below that we find the \texttt{starblade} algorithm. For it we can also state an uncertainty, which was calculated from one hundred samples of the approximate posterior, and it amounts to roughly ten percent in $\mathrm{cos}\theta$. This result mimics to some extent the outcome of the RMS test in the mock example. Overall \texttt{starblade} produces the most uncorrelated components, followed by the DAE and the highest correlation can be found in SExtractor.

\begin{table}
	\centering
	\caption{The cosine similarity between the diffuse and point-like component for all methods.}
	\label{tab:correlation_table}
	\begin{tabular}{|l|l|} 
		\hline
		Method & cosine similarity\\
		\hline
		\texttt{starblade}  & $0.0059 \pm 0.0005$\\
		SExtractor  & $0.094$\\
		DAE & $0.0082$\\
		\hline
	\end{tabular}
\end{table}

Another interesting question is how similar the \texttt{starblade} results are to the other methods. SExtractor performs reasonably well in most cases, so we do not want to diverge too strongly from its results, at least for the point-like component.  The results of the similarity between the different methods and with the data can be seen in Tab. \ref{tab:methods_correlation_table}. The first entry shows the cosine similarity between the point-like component and the data. Here the most dominant contributions origin from the brightest sources, therefore the high score. A larger similarity can be observed to the SExtractor point sources, which tells us that, at least for the bright sources, the methods behave highly similarly. The difference should be due to the deviations in the fainter sources and the remaining large-scale structures in SExtractor. The results for the DAE are very close to \texttt{starblade}, which we already observed in the power spectra. 

For the diffuse components, the picture is slightly different. The similarity to the data is low, as all bright sources are missing. To SExtractor it is significantly higher, as it correctly picks up the largest scales, which are responsible for the highest  contribution to the similarity, but they are still quite un-similar. Compared to the DAE, the similarity is very high, but significantly lower than in the point-like component. This again reflects our observations from the power spectra concerning the smaller and smallest scales, which diverge to some extent. Our estimated error for the point source similarities is one order of magnitude smaller compared to the diffuse component. This should be due to the robustness of the separation of the brightest sources, which impact the similarity the most. 

\begin{table}
	\centering
	\caption{The cosine similarity of the diffuse and point-like component between \texttt{starblade} and other methods and data.}
	\label{tab:methods_correlation_table}
	\begin{tabular}{|l|l|} 
		\hline
		component of \texttt{starblade} with & cosine similarity\\
		\hline
		point sources and data  & $0.9518 \pm 0.0003$\\
		point sources and SExtractor  & $0.9775 \pm 0.0003$\\
		point sources  and DAE  & $0.9983 \pm 0.0002$\\
		diffuse and data  & $0.3124 \pm 0.0007$\\
		diffuse  and SExtractor  & $0.790 \pm 0.002$\\
		diffuse  and DAE  & $0.983 \pm 0.002$\\
		\hline
	\end{tabular}
\end{table}

As previously mentioned we do not have access to any kind of ground truth in this real data case, so the judgment has to be subjective. First of all we do obtain satisfying results with the \texttt{starblade} algorithm also on real data. We separate point sources and diffuse emission also in the presence of point-spread functions and noise. Any kind of artifacts which are introduced by the measurement process and incompatible with diffuse emission are, as expected, attributed to point-like emission. Aiming at a reasonable separation of point-like and diffuse emission, SExtractor does not provide a useful result. We should note that the background estimation of SExtractor was also not designed for this particular  purpose. The DAE generalizes to some extent, especially in removing point-like emission, but lacks precision in removing point-sources from diffuse emission. Overall we would judge that  \texttt{starblade} provides the separation with highest accuracy also in this real data application, at least given all applied metrics.



\section{Conclusion}
\label{sec:conclusion}

We derived the \texttt{starblade} algorithm, which is capable of separating point-like from diffuse emission. It enforces positivity of all components and the correlation structure of the diffuse component is inferred as well. The only free parameters correspond to assumptions of the underlying point source distribution, for which physically motivated choices are available. As we perform a variational approximation to the true posterior, one has access to uncertainties on the separation itself, as well as all derived quantities.

We validate the implementation of the algorithm in an example with data generated according to the model, where it performs better in terms of the logarithmic root mean squared error than the background estimation of SExtractor, a denoising auto-encoder trained on the same model. It also exhibits more robustness in choice of the hyper-parameters than the MAP solution. 

Applying the algorithm to a data set of the $\mathrm{M}100$ galaxy obtained by HST provides satisfying results. The components are clearly separated visually and this impression is confirmed in the individual power spectra of the separated components. Of all applied methods, \texttt{starblade} provides the most uncorrelated components. A comparison between the results show a high similarity to the point-source result of SExtractor. 

The results of the \texttt{starblade} algorithms can be used in further analysis to build catalogs or to study extended, correlated structures. Through the samples, the uncertainty of the separation can be fully propagated to the science result at the end by performing all calculations for the samples, averaging the sample results and evaluate their variance. By this approach the full uncertainty is taken into account, including large scale effects from the diffuse component.

This method can also be used as an internal step within a larger inference framework, which solves the full reconstruction problem with all instrumental effects. By providing a good estimate of the separation of the components it can speed up the computations. Details on this are outlined in Appendix \ref{ap:lager_picture}.

We believe that the \texttt{starblade} algorithm can be used in a large variety of applications and we provide an open source application of it at \url{https://gitlab.mpcdf.mpg.de/ift/starblade}.
\section*{Acknowledgments}
We acknowledge Philipp Arras, Fabrizia Guglielmetti, Reimar Leike, and Martin Reinecke for fruitful discussions and comments on the manuscript.

\bibliographystyle{mnras}
\bibliography{citations} 

\appendix
\section{The larger picture}
\label{ap:lager_picture}
At this point let us briefly sketch how the here presented \texttt{starblade} algorithm can be included as an intermediate step in a grander reconstruction scheme, which is required for a large number of real world problems. In these the assumption of a noise-free and complete data set without any instrument effects is not justified and those complications have to be taken into account. For example in the case of radio interferometry the data are affected by Gaussian noise and only individual Fourier components of the sky brightness are measured. In case of photon count observations one has to deal with masking of some areas in the image, Poissonian shot noise, and point-spread functions.

For both measurements, radio interferometry and photon counting, it makes sense to assume the same underlying sky model, i. e. the superposition of diffuse emission and point sources $I_{sky} \equiv e^s + e^u$. In order to perform the reconstruction of both components one can set up a Bayesian inference scheme to obtain posterior estimates. This is straight forwardly done by using Bayes' theorem
\begin{align}
\mathcal{P}(s.u\vert d) = \frac{\mathcal{P}(d\vert s,u) \mathcal{P}(s)\mathcal{P}(u)}{\mathcal{P}(d)} \text{,}
\end{align}
and then usually some kind of approximation is applied, such as maximum posterior or a variational approach. In both approaches some target functional is minimized to get the final reconstruction. However, as we do have two sky components, both fully capable of explaining any kind of sky brightness, the separation during the reconstruction is delicate. The main driving force of the minimization is the gradient of the likelihood, which can be orders of magnitude stronger than any prior contribution. After the first few iteration steps the likelihood is relatively satisfied, but the separation of both components is more or less arbitrary as the weak prior could not establish its influence yet. The prior pushes the sky components into the right direction, but any change in one individual component has to fight against the gigantic potential walls of the negative log-likelihood and the separation process is numerically exhausting with only minimal improvements in each step. At this point the here presented method can be used as an intermediate procedure to sort out the component separation while keeping the current likelihood constant. This corresponds to an optimization along a sub-manifold in the Hilbert space of all sky images which keeps the likelihood constant and the prior terms can act freely, dramatically speeding up the overall reconstruction. Conceptually we expand the joint probability of all variables by introducing a delta distribution which decouples the prior quantities from the likelihood:
\begin{align}
\mathcal{P}(d,s,u) =& \int \mathcal{D} I_{sky}\: \mathcal{P}(d\vert I_{sky})\: \mathcal{P}( I_{sky}\vert s,u) \:\mathcal{P}(s)\: \mathcal{P}(u) \\
=&  \int \mathcal{D} I_{sky}\: \mathcal{P}(d\vert I_{sky})\:\underbrace{ \delta(I_{sky} - e^u -e^s) \:\mathcal{P}(s)\: \mathcal{P}(u)}_{\mathcal{P}(I_{sky}, s, u)}
\end{align}
The resulting sub-problem restricted by $\mathcal{P}(I_{sky}, s, u)$ is exactly the case discussed in this paper, which we can solve efficiently. After the components are separated properly one can go back to the full problem to continue the overall reconstruction by allowing the data to ask for different skies.

\section{Denoising convolutional auto-encoder}
\label{ap:DAE}
The  denoising auto-encoder (DAE) \citep{DAE} is a powerful tool to denoise images. It can be adapted for various kinds of noise effects, such as Gaussian noise or salt-and-pepper noise. It is also applicable to restore distorted images, such as compression artifacts from images. It consists of an encoding part, a latent layer and a decoding part. The net is trained on artificially corrupted images, for which the ground-truth is known. The input is the noisy image and the output should be the denoised version. 

The model described in this paper can be interpreted as diffuse emission which is corrupted by inverse-gamma distributed noise, so the DAE should be an excellent way to separate the diffuse emission from the point-sources. In order to train the auto-encoder we generate a data set according to the model described in this paper for a given set of parameters. In this case we do have access to the ground truth in form of the diffuse component. 

We took an off-the-shelf \texttt{Keras} \citep{keras} implementation of the denoising auto-encoder and adjusted it according to this problem. The encoder consists of two convolutional layers with a kernel size of $3$. The first layer consists of $32$ and the second layer of $16$ units. Compared to standard auto-encoders we clipped the fully connected latent layer, as we do not expect large-scale features to impact the separation too much. This significantly reduces the number of trainable parameters and we stronger rely on the locality of the problem. The decoder therefore directly follows with adjoint convolutions symmetrically to the encoder. Additionally we apply drop-out during the training to avoid over-fitting and force the net to generalize better. The drop-out rate is $0.1$ and it is performed after each convolutional layer. All convolutional layers have \texttt{relu} activation functions \citep{relu}. After the last convolution layer a sigmoid function is applied to follow the philosophy of the separation field as described in the paper. This proposed separation is merged with the input layer again by multiplication to generate the final output of the auto-encoder. This way the auto-encoder does not have to propagate information on the absolute image values and it can concentrate on more abstract concepts. We hope that this speeds up the learning of the parameters, as well as increasing the overall performance.

Using a mean squared error loss function to learn the parameters is insufficient due to the dynamic range of the problem. Making small errors on bright sources is dis-proportionally punished, which leads to a resignation of the network. In this case it sets the diffuse component to constant zero and accepts a non-optimal solution. Any attempt to learn more abstract concepts are immediately punished and it reverts back to the constant state. 

To avoid this and to respect the exponential nature of this problem we instead train the network parameters by minimizing the logarithmic mean squared error loss. Here an error in the order of magnitude is only punished quadratically, which allows the auto-encoder to converge at non-trivial solutions.

The training data consisted of $2500$ images drawn from the model with the correct correlation structure and point-source statistics in a resolution of $128$ by $128$ pixels. The performance was validated on another $500$ images to detect signs of over-fitting during training. The loss was optimized using the Adam optimizer \citep{adam} with a batch size of $32$ images for $1000$ epochs with a \texttt{TensorFlow} $1.5$ \citep{tensorflow} back-end to \texttt{Keras} on a \texttt{Nvidia GTX 1080ti}, which took roughly two hours.

The performance of the DAE are discussed in Section \ref{sec:mock_example} together with the other methods.

\bsp	
\label{lastpage}
\end{document}